%% file: main.tex
\documentclass[lettersize,journal]{IEEEtran}
\usepackage{amsmath,amsfonts}
\usepackage{algorithm}
\usepackage{array}
\usepackage[caption=false,font=normalsize,labelfont=sf,textfont=sf]{subfig}
\usepackage{textcomp}
\usepackage{stfloats}
\usepackage{url}
\usepackage{verbatim}
\usepackage{graphicx}
\usepackage{cite}
\usepackage{tikz}
\usetikzlibrary{decorations.pathreplacing,calc}

\usepackage{algorithm,algpseudocode}
\usepackage{setspace}
\usepackage{caption}

\usepackage{xspace}
\usepackage{soul}
\newcommand{\sys}{\textsc{HADES}\xspace}
\newcommand{\descr}[1]{\smallskip \noindent \textbf{#1}}
\newcommand{\descrit}[1]{\smallskip \noindent}
\usepackage{subcaption}
\usepackage{multirow}
\usepackage{booktabs,tabularx}
\newcommand{\revised}[1]{{\color{black}{#1}}}
\usepackage{amsmath}
\usepackage{amsfonts}

\usepackage{amssymb}
\usepackage[hidelinks]{hyperref}
\usepackage{orcidlink}

\begin{document}

\author{
Erg\"un Batuhan Kaynak\,\orcidlink{0000-0002-3249-3343},
Kerem Bayramoglu\,\orcidlink{0009-0009-2850-9945},
and Sinem Sav\,\orcidlink{0000-0001-9096-8768}%

\thanks{All authors are with the Department of Computer Engineering, Bilkent University, Ankara, Turkey. Corresponding Author: Sinem Sav}%
\thanks{Ergün Batuhan Kaynak (e-mail: batuhan.kaynak@bilkent.edu.tr).}%
\thanks{Kerem Bayramoglu (e-mail: kerem.bayramoglu@cs.bilkent.edu.tr).}%
\thanks{Sinem Sav (e-mail: sinem.sav@cs.bilkent.edu.tr).}%
\thanks{Manuscript received XX XX, XXXX; revised XX XX, XXXX.}%
}

\title{HADES: Privacy-Preserving Federated Learning via Selective Feature Encryption and Hybrid Model Fusion}

\maketitle

\begin{abstract}
In this paper, we address the challenge of privacy-preserving training in federated learning (FL) by introducing a novel framework that selectively encrypts only the most privacy-sensitive features while leaving the remaining data and the corresponding model portion unencrypted. We propose \sys, a hybrid system that identifies and encrypts the most critical features, ensuring both privacy protection and computational efficiency. Unlike fully encrypted FL training pipelines, which suffer from high computational overhead, \sys integrates an encrypted and non-encrypted training pipeline via a fusion mechanism, enabling seamless interaction between encrypted and plaintext model representations. To achieve this, we use PCA to identify and encrypt the most privacy-sensitive features, which significantly reduces reconstruction attack success in FL. Building on this insight, we design a hybrid FL system that trains an end-to-end encrypted network via multiparty homomorphic encryption (MHE) on the selected features while simultaneously training a plaintext network on the remaining features. These two networks are then integrated using a fusion mechanism. We also introduce a general packing scheme that eliminates redundant rotations by considering the entire neural network architecture. Finally, we demonstrate that \sys matches the accuracy of vanilla FL while preserving privacy and achieving optimized runtime through selective encryption. 
\end{abstract}
\markboth{Journal of \LaTeX\ Class Files,~Vol.~14, No.~8, August~2021}%
{Shell \MakeLowercase{\textit{et al.}}: A Sample Article Using IEEEtran.cls for IEEE Journals}


\input{introduction}
\input{related}
\input{background}
\input{method}

\section{EXPERIMENTAL EVALUATION}
In this section, we experimentally evaluate \sys to address the following key questions:
\begin{itemize}
    \item \textbf{Q1:} Does PCA-based feature selection and encryption effectively protect against data reconstruction attacks in FL settings?
    \item \textbf{Q2:} Does \sys preserve the model utility?
    \revised{\item \textbf{Q3:} How does HADES compare against a baseline that trains solely on the PCA-selected feature subset,
    without employing dual network fusion?}
    \item \textbf{Q4:} How does \sys scale with the number of encrypted parameters ($|\mathcal{F}_{HE}|$ ), the number of clients in the FL, and the model complexity?
    \item \textbf{Q5:} What is the overall runtime performance of \sys?  
\end{itemize}

Finally, we note that comparing \sys to state-of-the-art encrypted FL frameworks is inherently challenging for several reasons: (i) The private FL literature generally falls into two categories: fully encrypted training and secure aggregation. \sys introduces a hybrid approach that applies model encryption and secure aggregation to only a subset of the parameter space, while still offering privacy guarantees for the unencrypted parameters. To the best of our knowledge, no prior work adopts this selective encryption strategy, leaving no directly comparable framework. (ii) The most closely related works—fully encrypted FL frameworks such as POSEIDON \cite{sav2021poseidon} and Hercules \cite{hercules}—encrypt all model parameters. As a result, under identical settings and packing schemes, \sys is expected to reduce runtime in proportion to the number of parameters it encrypts ($|\mathcal{F}_{HE}|$) Therefore, our scalability plots (Figure~\ref{fig:fhe_vs_time}) effectively illustrate this comparative advantage. (iii) \sys implements HE operations using the Python wrapper for OpenFHE \cite{OpenFHE}, whereas POSEIDON and Hercules are built on the Lattigo library. These frameworks prioritize HE-specific optimizations, while \sys focuses on enabling efficient selective encryption. 

In the following subsections, we first describe our experimental setup (Section~\ref{sec:experimentalSetup}), detailing the datasets, parameters, and implementation details. Then, each subsequent subsection systematically addresses the key questions outlined.
\subsection{Experimental Setup} \label{sec:experimentalSetup}

\par\noindent\textbf{Datasets.}
For our experiments, we use the Breast Cancer Wisconsin (Diagnostic) dataset \cite{breast_cancer_wisconsin_(diagnostic)_17}, containing 569 samples with 30 real-valued features per sample; the MNIST dataset \cite{lecun-mnisthandwrittendigit-2010}, consisting of 70,000 grayscale images of handwritten digits, each with 28×28 pixels (totaling 784 features per image); \revised{and the SVHN dataset \cite{Netzer2011ReadingDI}, which contains 99,289 32x32 RGB digit images (3072 features per image) cropped from real-world Street View scenes. We use a 70\%-30\% train--test split for all datasets (SVHN uses its predefined 73.78\%-26.21\% split).}

\par\noindent\textbf{Implementation.}
We implement our plaintext network and its operations from scratch using NumPy~\cite{harris2020array}, without relying on any existing machine learning frameworks. We opt for this to eliminate any computation changes that can happen due to internal library operations, and keep encrypted and plaintext network results close to each other for the same input (See Section \ref{Q2}). Network weights are initialized with \revised{Xavier} initialization, and bootstrapping is applied to the weights (of the encrypted sub-network) at the end of each weight update. Our loss function is implemented in a way that works with one-hot-encoded labels. This enables the generation of losses from multilabel data in an efficient manner as in~\cite{sav2021poseidon}. 

Our in-house HE operation framework is built as an extension for the OpenFHE-python library \cite{OpenFHE}. The framework includes efficient yet flexible alternating packing schemes \cite{sav2021poseidon} for matrices, convenient handling of padding, masking, and rotation of matrices, allowing mini-batch operations on multi-layer neural networks. Each ciphertext optionally holds information regarding its valid indices (i.e., non-zero values that are not ciphertext operation byproducts) and underlying padded representation to ease debugging. 

We evaluate the privacy of \sys using the success rate of the Improved Deep Leakage Gradient(iDLG)~\cite{zhao2020idlgimproveddeepleakage} attack. Following our previous argument regarding network implementations from scratch, we implemented iDLG using NumPy \cite{harris2020array} and L-BFGS\cite{R._2000} implementation from SciPy\cite{2020SciPy}. Since the gradients from the encrypted network cannot be recovered, all iDLG attack results come from the unencrypted plaintext network. \revised{We then evaluate the reconstruction rate of the recovered features using Root Mean Square Error (RMSE), Peak Signal-to-Noise Ratio (PSNR), Structural Similarity Index (SSIM) \cite{ssim} and Learned Perceptual Image Patch Similarity (LPIPS)\cite{lpips}.}


\par\noindent\textbf{System configuration.}
Each experiment model is trained for \revised{10 epochs} using Stochastic Gradient Descent (SGD) with a \revised{local} batch size $B = 1$ \revised{(i.e. a global batch size $ B_g = K$ for FL settings)} and number of clients $K = 10$, unless stated otherwise. A single-element batch size is chosen because it maximizes the success rate of the iDLG attack, providing a rigorous evaluation of \sys under challenging conditions. \revised{Furthermore, it allows us to more clearly see the limitations of not using \sys sub-network strategy, since such a network would only support a very small mini-batch, and that is if the network dimensions are kept smaller. In Section \ref{Q5}, we show how \sys seamlessly allows for fitting larger mini-batches in a single ciphertext.} In decentralized settings, we define an epoch as one complete pass through the entire dataset. In multiparty setups, each client processes $\frac{|D|}{n_k \cdot B}$ global iterations \revised{(GI)} to go through their entire data, where $|D|$ is the total sample size of the dataset, and $n_k$ is the number of samples for client $k$. The network configurations change based on different experiments, and they are detailed in their respective sections. In each experiment, both plaintext and encrypted sub-networks use the same number of layers and layer dimensions. The score-level fusion trade-off parameter $\alpha$ is set to 0.5, ensuring equal contribution from both sub-networks during training. The cyclotomic ring size for CKKS is set to $\mathcal{N} = 2^{13}$, yielding a ciphertext vector in $\mathbb{C}^{\mathcal{N}/2}$ with $2^{12}$ slots. We use a 64-bit precision and set the number of levels to $L=6$.

\subsection{Q1: Does PCA-based feature selection and encryption effectively protect against data reconstruction in FL settings?}\label{sec:q1}

To evaluate the effectiveness of our feature selection approach in preserving privacy, we conducted preliminary experiments using the \textit{Improved Deep Leakage from Gradients (iDLG)}~\cite{zhao2020idlgimproveddeepleakage} attack framework against multiple datasets. In iDLG, the recovery process begins by initializing a random input, $\mathbf{X}$, and iteratively updating it to minimize the difference between the gradients of the hypothetical input and the leaked gradients. This optimization minimizes the gradient mismatch, defined as $\nabla\mathcal{L} = \|\nabla{\mathbf{X}} - \nabla{W_{P}^{k, t}}\|^2$, where $\nabla{W_{P}^{k, t}}$ represents the gradients of non-encrypted features shared during training. By iteratively refining $\mathbf{X}$, we ensure alignment between the hypothetical and leaked gradients.

Our proposed defense against such gradient-based attacks relies on selectively encrypting a subset of the feature space. We select these features based on the \revised{cumulative} explained variance as calculated by PCA. Table \ref{tab:pca_baseline_comparison} \revised{reports, for each dataset, how much variance is retained as the number of preserved principal components increases, and pairs this with test accuracy for (i) a \textbf{Baseline} model trained using only the encrypted PCA features and (ii) \textbf{\sys}, which additionally leverages the remaining (unencrypted) features through fusion.

As expected, a significant portion of the variance can be explained by a relatively small subset of the original feature space. This is particularly evident in the BCD dataset, and we argue that this characteristic is a primary reason why the performances of the baseline and \sys are both similar and erratic. However, explained variance alone does not guarantee predictive performance: for MNIST and SVHN, utilizing very few encrypted components yields poor baseline accuracy despite non-trivial retained variance, whereas \sys maintains substantially higher accuracy by exploiting complementary information from the non-encrypted features. Encrypting only these principal directions preserves most of the privacy-sensitive signals while sharply reducing the plaintext surface available to an attacker. Thanks to the \sys fusion approach, the remaining features are also utilized for training, ensuring their information is not lost.

}

\input{Tables/pca_variance_table}

 \revised{Table~\ref{tab:recon_stats} reports reconstruction quality under the \emph{random-initialization} gradient inversion attack as a function of the number of encrypted PCA features, $|\mathcal{F}_{HE}|$. Reconstruction fidelity is summarized using RMSE (lower is better), PSNR in dB (higher is better), SSIM in $[0,1]$ (higher is better), and LPIPS (lower is better). As expected, when $|\mathcal{F}_{HE}|=0$ (no encryption; \sys disabled), the attacker achieves perfect reconstruction across datasets (RMSE $=0$, PSNR $=\infty$, SSIM $=1$, LPIPS $=0$). Increasing $|\mathcal{F}_{HE}|$ steadily degrades reconstruction: on MNIST, moving from $16$ to $256$ encrypted features reduces SSIM from $0.48$ to $0.11$ and increases LPIPS from $0.17$ to $0.49$, while PSNR drops from $13.48$\,dB to $8.92$\,dB; on SVHN, SSIM falls from $0.53$ (at $16$) to $0.04$ (at $256$) and LPIPS rises from $0.16$ to $0.53$, indicating that class-defining structure is largely lost. Importantly, a \emph{moderate} encryption budget already approaches the dataset-specific worst case $|\mathcal{F}_{HE}|=|F|-1$: for SVHN, SSIM at $256$ is $0.04$ versus $0.01$ at $|F|-1$, and for MNIST it is $0.11$ versus $0.04$, suggesting diminishing returns beyond this point. Overall, these results support our claim that selectively encrypting a modest subset of features is sufficient to substantially suppress reconstruction quality, approaching the $|F|-1$ regime without encrypting nearly all features.
}

\input{Tables/recon_attack_table}

\revised{To complement the quantitative metrics in Table~\ref{tab:recon_stats}, Figure~\ref{fig:recon_fusion_256} provides qualitative reconstructions produced by the same attack. Following our previous argument, we pick $|\mathcal{F}_{HE}|=256$ to present our reconstruction examples, which already approach the near-worst-case reconstruction on MNIST and SVHN. Across different inputs, the reconstructed images exhibit no perceptible similarity to the original content. This collapse toward similar reconstructions is consistent with the substantial loss of discriminative information when the selected features are encrypted: the attacker is left to fit secondary features with virtually no identifying information.}

\input{Figures/recon_images}

These findings highlight the practical effectiveness of our feature selection mechanism. By strategically selecting and protecting features, significant privacy improvements can be achieved under realistic attack conditions, effectively reducing the risk of data leakage without compromising the utility of FL systems (see Section~\ref{Q2}). Moreover, the observed disparities in attack efficacy between theoretical and practical scenarios highlight the critical role of adversary knowledge and initialization conditions in data reconstruction attacks.

\subsection{Q2: Does \sys preserve the model utility?}
\label{Q2}

For all three datasets, we conduct 4 experiments and report their test accuracies in Table \ref{tab:exp2_results}. These tests collectively evaluate and compare the results of (a) centralized (CE) vs federated learning (FL) training setup, (b) using a vanilla or approximated sigmoid (approx) activation function, and (c) using a single network ($\mathcal{F}_{HE} = \emptyset$) or the proposed fusion network for a given $\mathcal{F}_{HE}$. For experimental purposes, we simulate the encryption process using cleartext in these experiments. Due to the difference in feature sizes and numeric instability in certain cases, these experiments are run using a certain hyperparameter configuration for each dataset. We chose these hyperparameters via brief manual tuning on the training data. Specifically, for BCD we use a hidden layer of 16 units, an encrypted‐feature packing factor $|\mathcal{F}_{HE}|=16$. For MNIST, we increase the hidden layer to 64 units and $|\mathcal{F}_{HE}|$ to 256. \revised{For SVHN, we use two hidden layers with 64 units each and keep $|\mathcal{F}_{HE}|$ at 256. A learning rate of 0.1 with 0.9 nesterov momentum was used to train 10 federated clients for 10 epochs for all datasets. We do not optimize these parameters across datasets or configurations to better show the consistent dynamics of \sys across datasets. These $|\mathcal{F}_{HE}|$ values account for respectively 2x, 4x, and 16x reduction in encrypted parameters compared to their non-\sys counterparts. 

Importantly, we use local batch size $B = 1$ and global batch size $B_g = 10$ in this experiment. As a result, the centralized (CE) configurations benefit from more stable updates due to aggregation over a larger batch, whereas the federated (FL) setting exposes per-client gradients computed from a single input sample, making the updates significantly noisier. We intentionally adopt this regime because gradient inversion attacks such as iDLG are most favorable when an iteration’s gradients are produced by a single input (i.e., $B=1$), yielding the easiest reconstruction condition for the attacker.}

\input{Tables/exp2_table}


\revised{Across all settings and datasets, performance differences remain modest, indicating that our approximation and fusion schemes preserve predictive utility. The sigmoid approximation closely matches the results obtained with exact activations, and the fusion network does not introduce a meaningful degradation. Notably, \sys achieves the strongest gains on SVHN, improving accuracy from 63.1\% (CE-$\mathcal{F}{HE}{=}0$) to 70.4\%, which suggests that the hybrid design can recover substantial task-relevant signal when the input space is high-dimensional and heterogeneous. While CE-$\mathcal{F}{HE}$ and FL-$\mathcal{F}_{HE}$ include more parameters due to the fusion setup, their encrypted-feature selection is random; thus, they serve as a conservative baseline for the fusion architecture. In contrast, \sys benefits from a guided split that places salient components in the encrypted branch and leaves complementary information to the plaintext branch, enabling the fused predictor to better exploit both protected and residual structure.}

\revised{
\subsection{Q3: How does \sys compare against a baseline that trains solely on the PCA-selected feature subset, without employing dual network fusion?}
\label{Q3}

Given our selective-encryption design, a natural question is whether, once PCA identifies the most informative or privacy-sensitive features, one could train a single network using only this selected subset and discard the remaining features. To isolate the effect of PCA-selected features from the benefits introduced by \sys's hybrid architecture, we include a single-network ablation in which only the PCA-selected components are used to train the model. This ablation removes the fusion mechanism, allowing a direct comparison of a baseline architecture against \sys's encrypted model using the same feature subset. Both architectures are agnostic to the choice of activation approximation. To avoid confounding effects from approximation quality, we simulate this ablation with non-approximated activations for both models. This is also the fairest comparison point, since tuning approximation degrees or ranges per model can grant an advantage to one architecture over the other.

\begin{figure*}[ht!]
    \centering
    \includegraphics[width=0.9\linewidth]{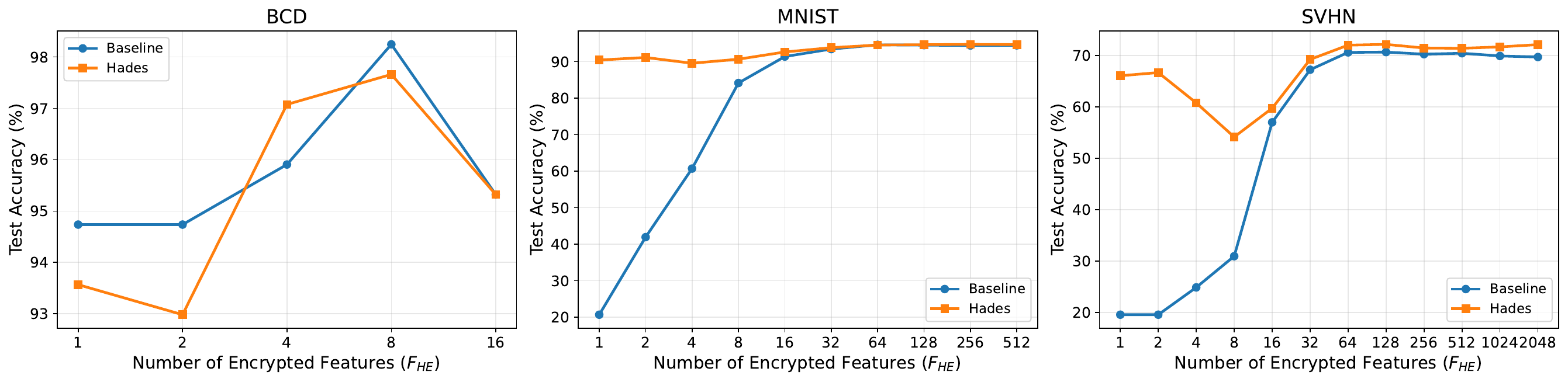}
    \caption{Test accuracy (\%) versus the number of encrypted PCA-selected features ($|\mathcal{F}_{HE}|$) for BCD, MNIST, and SVHN. \emph{Baseline} trains a single network using only the selected subspace, while \sys applies score-level fusion between an encrypted branch on $\mathcal{F}_{HE}$ and a plaintext branch.}

    \label{fig:pca_baseline_comparison_graph}
    \vspace{-1em}
\end{figure*}

As shown in Table~\ref{tab:pca_baseline_comparison} and Figure~\ref{fig:pca_baseline_comparison_graph}, training solely on the PCA-derived subspace leads to a clear accuracy drop across MNIST and SVHN, indicating that selecting only the “most important” components is often insufficient and discards task-relevant information. The BCD results are less consistent because even a single principal component captures the vast majority of the variance, reflecting the limited and highly specific feature set of this dataset. As a result, PCA-based feature pruning is comparatively less informative, and relying on it alone leads to a reduction in model utility. In contrast, \sys recovers utility through score-level fusion: the plaintext branch preserves fine-grained patterns that the selected components omit, while the encrypted branch protects the most sensitive components. This ablation highlights that the gains in accuracy stem from the fused dual-network design rather than from PCA alone. Across datasets, accuracy generally improves with larger $|\mathcal{F}_{HE}|$, and \sys is consistently more robust at small encrypted-feature budgets, with both methods converging as more features are included.


}

\subsection{Q4: \revised{How does \sys scale with the number of encrypted parameters ($|\mathcal{F}_{HE}|$ ), the number of clients in the FL, and the model complexity?}}
\label{Q4}

\begin{figure*}[!t]
\centering
\subfloat[Width scaling for a single (or no) hidden layer.]{
  \includegraphics[width=0.4\textwidth]{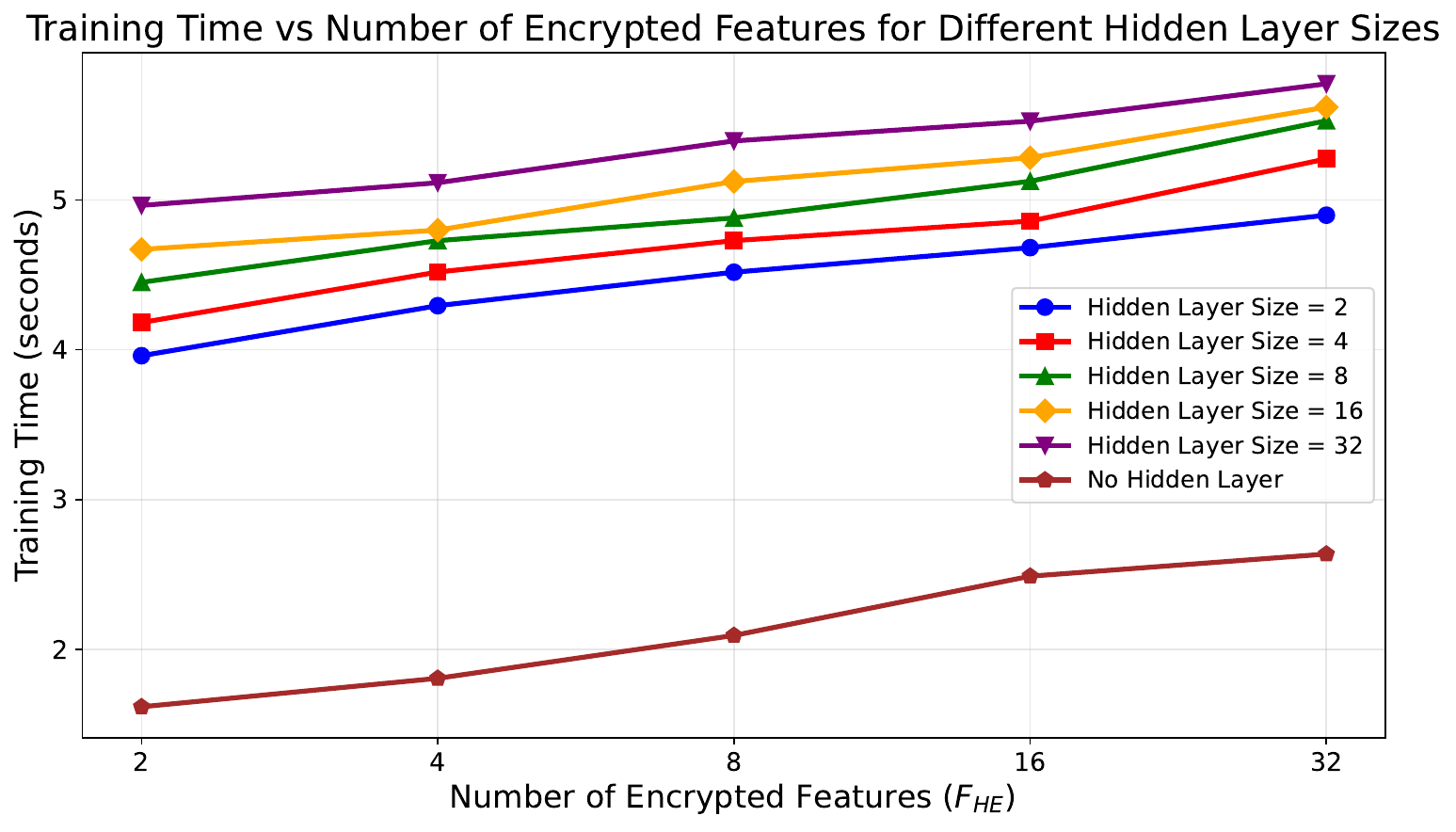}
  \label{fig:fhe_vs_time_shallow}
}
\subfloat[Depth scaling for a single hidden layer.]{
  \includegraphics[width=0.4\textwidth]{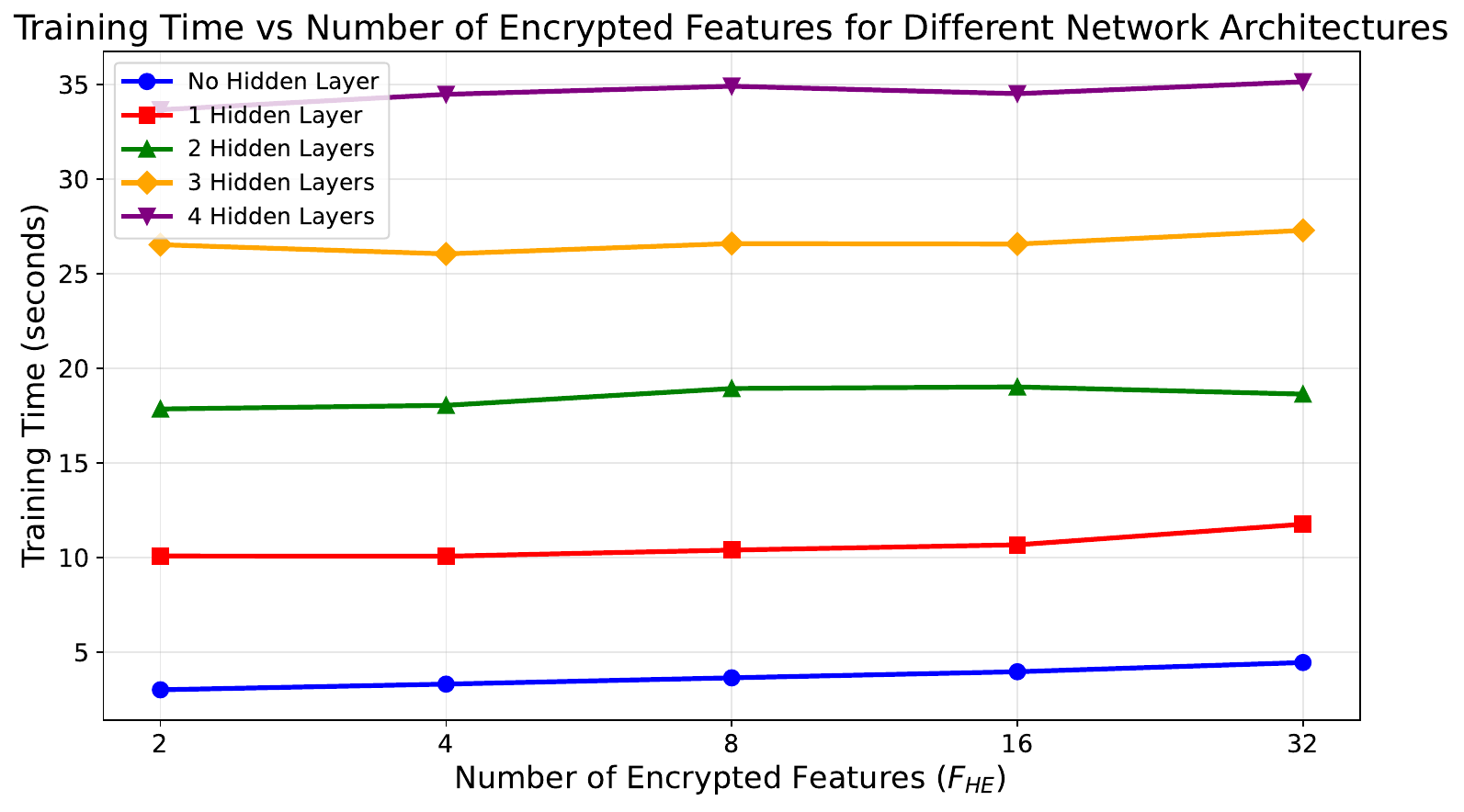}
  \label{fig:nlayer_vs_time}
}
\caption{Training time as a function of the encrypted-feature set ($F_{HE}$) for models with (a) 0, 2, 4, 8, 16, and 32 hidden units and (b) increasing number of layers with 16 hidden units each. Runtime increases linearly with $F_{HE}$ for every depth: wider networks start with higher runtimes but scale at a similar rate, whereas the no-hidden-layer configuration is consistently the fastest with the same upward trend.}
\label{fig:fhe_vs_time}
\vspace{-1em}
\end{figure*}

The experimental results in Figure~\ref{fig:fhe_vs_time} report the training time using a synthesized dataset. To account for variability, each experiment was repeated 10 times across 10 distinct data points, and the results were averaged. A synthesized dataset with 32 features was used in this experiment, where $|\mathcal{F}_{HE}|=32$ denotes training under full encryption, that is, with all features treated as private. These results allow for an analysis of how \sys scales with respect to the number of homomorphically encrypted features ($|\mathcal{F}_{HE}|$) and the complexity of the encrypted sub-network (specifically, its hidden layer size).

A primary observation, which was the core motivation for \sys, is the direct correlation between the number of encrypted features and the computational overhead. As $|\mathcal{F}_{HE}|$ decreases, training time also decreases. Based on network configuration, this improvement can be up to $28\%$. The complexity of the neural network architecture operating on encrypted data also significantly impacts performance. With $|\mathcal{F}_{HE}|$ fixed at certain values, the introduction of a hidden layer into the encrypted sub-network, and the subsequent increase in its size leads to a substantial rise in training times. Wider hidden layers consistently incur a higher absolute cost, but the incremental penalty of doubling $F_{HE}$ remains comparable, with each step adding on the order of a few hundred milliseconds. The "No Hidden Layer" baseline is substantially faster overall but follows the same upward trajectory, indicating that ciphertext packing, not hidden-layer computation, dominates runtime at large $F_{HE}$. Collectively, the plot demonstrates that training time scales approximately linearly with the number of encrypted features. For deeper networks, we observe diminishing returns from tuning $|\mathcal{F}_{HE}|$. Even then, we still maintain $\sim7\%$ decrease in runtime in worst-case configurations. 

This trend is also consistently observed in an increasing number of clients. This is expected, as the improvements introduced by \sys primarily stem from local training steps. The overall FL procedure remains unchanged, aside from the negligible communication overhead introduced by transferring unencrypted gradients.

We also provide the number of costly operations per forward and backward pass of the network in Table \ref{tab:ops_counts}. Each additional encrypted layer introduces a fixed computational overhead: 4 extra rotations, 1 plaintext–ciphertext multiplication (MultPT), and 3 ciphertext–ciphertext multiplications (MultCT). Since rotations dominate latency calculations, overall compute grows based on the number of rotations. As previously argued, the fusion point of the network can be in any intermediate layer. This decision should be given based on the computational considerations presented here, and an additional security analysis conducted for that specific network architecture.

\begin{table}[h]
\centering
\small
\caption{Operation counts for forward and backward passes for an $n$ layer network, per client.}
\label{tab:ops_counts}
\begin{tabular}{lccc}
\toprule
 & Rotation & MultPT & MultCT \\
\midrule
\multicolumn{4}{l}{\textbf{Forward}} \\
Count & $2n - 1$ & $2n$ & $n$ \\
\midrule
\multicolumn{4}{l}{\textbf{Backward}} \\
Count & $2n - 1$ & $2n + 1$ & $2n + 1$ \\
\midrule
\multicolumn{4}{l}{\textbf{Total}} \\
Count & $4n - 2$ & $4n + 1$ & $3n + 1$ \\
\bottomrule
\end{tabular}
\end{table}

In summary, the scalability of \sys is influenced by the number of sensitive features ($\mathcal{F}_{HE}$) and the architectural complexity of the encrypted sub-network. The system exhibits proficient scaling with an increasing number of clients in terms of per-client, per-data point computational efficiency. While more complex HE networks inherently increase the baseline computational cost, the scaling trend with an increased number of clients remains largely positive or neutral, showing the viability of \sys in distributed FL environments.

\input{Tables/gi_per_dataset}

\subsection{Q5: What is overall the runtime performance of \sys?}
\label{Q5}

\revised{In this section, we report runtime measurements across multiple datasets and configurations. We introduce \textit{GI-Load} as the number of global iterations (GI) required by a configuration to process a fixed number of samples. In our setup, GI-Load scales approximately inversely with the batch size $B$: for example, a configuration with $B=8$ processes 80 samples in 10 iterations, whereas $B=16$ processes the same 80 samples in 5 iterations, resulting in half the GI-Load.} Table~\ref{tab:gi_dataset} shows that for each dataset, increasing $B$ incurs a slight cost on One-GI time. This is due to the additional rotation operation needed in each forward and backward pass to process the mini-batch. However, \revised{this added latency is amortized by the reduction in GI-Load, which reduces the number of bootstraps, rotations, and communication rounds over the full training run. For comparability across settings, we fix the iteration count to 100 when GI-Load $=16X$ and scale it linearly for smaller loads as $I_{kX}=100\cdot k/16$. Finally, for a fixed GI-Load (i.e., fixed $B$), One-GI is primarily influenced by $|\mathcal{F}_{HE}|$, which changes ciphertext utilization and the per-iteration homomorphic cost.}

When using a batch size of $B = 1$, application-level $P$-parallelization combined with local gradient accumulation can be employed to effectively simulate a batch size of $B = P$. Such an approach would still preserve the privacy guarantees of this study and reduce One-GI by two means: hardware parallelism and fewer homomorphic operations for single-cipher mini-batches. However, our focus on this study is to show how \sys can increase single-cipher utilization without relying on hardware-level parallelization.

\section{Discussion}

An important design consideration in \sys is minimizing the number of slots consumed by the feature dimension, as this enables more efficient use of the ciphertext space for operations such as expanding hidden layer dimensions or increasing mini-batch size. To better demonstrate this emergent property, we adopt a packing strategy that encodes the entire mini-batch into a single ciphertext. This contrasts with approaches like POSEIDON~\cite{sav2021poseidon} and Hercules~\cite{hercules}, which either parallelize over individual examples within the mini-batch or process them sequentially before averaging. By leveraging a single-ciphertext strategy, \sys significantly reduces memory usage and CPU core requirements during inference and training. The parallel approach allows for further optimizations, such as omitting rotations in the forward and backward pass of the first layer using duplication of the input data, while packing the mini-batch in the same ciphertext requires two additional rotation calls.

\revised{A key advantage of \sys is that its value proposition follows naturally from the cost structure of SIMD HE. Reducing the encrypted feature dimension decreases ciphertext slot consumption, which typically lowers the cost of downstream encrypted linear algebra (e.g., encrypted matrix–vector products and rotations) and therefore reduces end-to-end training time. On the other hand, the plaintext support network provides extra representational capacity "for no cost" in HE terms: its computation is inexpensive relative to ciphertext arithmetic, yet it increases the overall number of parameters and thus capacity, utility, and flexibility.}

\revised{Our empirical evaluation utilizes \sys with fully connected networks. However, \sys is \emph{not} tied to a specific neural architecture: it is a protocol-level design that composes two subnetworks (an encrypted branch operating on $\mathcal{F}_{HE}$ and a plaintext branch operating on $\mathcal{F}\setminus \mathcal{F}_{HE}$) and combines their predictions (or intermediate representations) via a fusion operator. Consequently, \sys can be integrated with any neural network backend as long as it supports a fusion point. Thus, \sys is complementary to existing encrypted FL methods rather than a replacement. \sys specifies \emph{what} portion of the feature space is encrypted and \emph{how} encrypted and plaintext computations are combined.
}

\sys introduces several mechanisms to balance privacy and utility that can be tuned based on circumstances. Our selective encryption scheme depends on a feature selection algorithm. This algorithm can be improved by future work to provide more security even with a smaller subset of encrypted features. Moreover, based on the privacy needs and the domain, fewer features can be encrypted to improve the performance significantly. The choice of layer for fusion allows for tuning the trade-off between accuracy, training time and privacy. In this study, we opt for score-level fusion with equal contribution from both plaintext and encrypted models ($\alpha = 0.5$). Weighted averaging parameter $\alpha$ can be tuned based on domain if, say, most private features are also more important for model performance or vice versa. Additionally, intermediate fusion techniques, such as concatenation or addition of hidden layer representations, offer further flexibility in optimizing the model's performance, while reducing the performance overhead. By employing selective HE, we optimize communication efficiency while preserving privacy. The bulk of the communication overhead is from encrypted gradients. Significantly less traffic is created as $|\mathcal{F}_{HE}|$ is tuned to be lower. Also, the smaller GI-Load of models under \sys results in fewer communication rounds, further reducing overall communication overhead.

We adopt PCA as a proof-of-concept feature selection method due to its computational efficiency, data-agnostic nature, and ease of reproducibility; importantly, \sys is fully agnostic to the choice of feature selection technique. We note that PCA's variance-maximization objective does not inherently provide privacy guarantees. Despite this, our results (Table~\ref{tab:recon_stats} and Figure~\ref{fig:recon_fusion_256}) demonstrate that encrypting \revised{a small amount of principal components} is sufficient to drastically reduce \revised{reconstruction success}, while preserving model accuracy (Section~\ref{Q2}). Exploring more targeted feature selection strategies—such as privacy-aware information ranking, adversarial filtering, or task-specific neural saliency—remains a promising avenue for future work, particularly for enhancing protection of privacy-sensitive attributes. This is especially pertinent for commonly used datasets like MNIST, where a small set of highly salient features dominate classification, leaving much of the remaining feature space underexplored in terms of privacy impact.


For our privacy-related analyses and experiments (see Section~\ref{Q2}), we adopt the strongest known attack setting by using a batch size of $B = 1$, where the adversary has access to gradients from individual samples. In contrast, when $B > 1$, the attacker only observes an average gradient across multiple inputs and must disentangle them to recover specific examples—a significantly more challenging task. Notably, our results reveal that reducing the feature space enables the use of larger batch sizes within a single ciphertext. This emergent property not only improves computational efficiency but also enhances security, indicating that \sys becomes increasingly robust in practical deployments.

\section{Conclusion}

In this paper, we introduced \sys, a novel privacy-preserving FL framework that selectively encrypts only the most privacy-sensitive features while processing the remaining data in plaintext. In contrast to secure aggregation techniques, \sys preserves encryption over the protected network components for the entire duration of training. By leveraging PCA for feature selection and integrating encrypted and plaintext sub-models through a fusion mechanism, \sys strikes an effective balance between privacy and computational efficiency. Our results demonstrate that \sys significantly mitigates reconstruction attacks while maintaining model accuracy comparable to standard FL. Moreover, the selective encryption approach reduces the computational and communication overhead associated with fully encrypted training pipelines. To the best of our knowledge, \sys is the first system to perform hybrid model fusion between encrypted and plaintext components in FL, offering a practical solution for privacy-sensitive collaborative learning tasks. 

\bibliographystyle{IEEEtran}
\bibliography{ref}

\IEEEpubidadjcol

\newpage
\input{supplementary}

\end{document}

%% file: introduction.tex
\section{Introduction}

In the era of big data and machine learning, data privacy has emerged as a critical challenge. As the demand for highly accurate models grows, so does the need for vast datasets. However, this pursuit often involves handling sensitive information, bringing significant privacy risks. Thus, ensuring ethical data practices while fostering innovation is imperative. To address the growing need for large-scale data while enhancing privacy, federated learning (FL)~\cite{Konency2016fed,federatedLearning1} has emerged as a collective machine learning solution. FL enables model training across vast, distributed datasets stored on local premises, ensuring data remains on its source rather than being centrally collected. Rather than transferring data to a central model, FL brings the model to the clients via an aggregation server. Clients then update the model using their local data, and these local updates from multiple parties are aggregated to construct the global model through iterations. However, despite these advantages, numerous attacks have shown that sensitive information remains at risk. Sharing intermediate model updates—whether among parties or with the server—can create vulnerabilities such as input extraction~\cite{bagdasaryan2020backdoor,Hitaj2017,Wang2019} and membership inference attacks~\cite{Melis2019,Nasr2019,zhang2020gan}.

To address privacy risks in FL, current research frequently leverages differential privacy (DP) and homomorphic encryption (HE).
Several studies incorporate DP to protect input data or intermediate value exchanges between server and FL parties~\cite{shokri2015privacy,McMahan2018,truex2020ldp}. Although these techniques help limit privacy attacks, they often compromise data utility and model performance. In addition, several works demonstrate that DP might be insufficient in several FL frameworks~\cite{you2024local}. HE-based solutions, on the other hand, generally follow one of two strategies: (i) secure aggregation of client inputs during model updates~\cite{mansouri2023sok,hosseini2021secure,zhang2020batchcrypt,bonawitz2017practical}, or (ii) encrypting the entire FL training process to maintain end-to-end privacy~\cite{sav2021poseidon, Sav2022PrivacyPreservingFR,hercules,9833648}. While the first approach lacks \revised{thorough} privacy protection, as the decrypted model on the client side remains susceptible to privacy attacks, the second approach significantly increases computational overhead due to HE operations. For example, in a fully encrypted FL pipeline—where intermediate values, local models, and the global model remain encrypted throughout execution—training can take from several hours to multiple days, depending on the network architecture~\cite{sav2021poseidon}.

To overcome the limitations of HE-based FL, we propose \sys, a novel framework that leverages multiparty homomorphic encryption (MHE) to enhance privacy-preserving neural network training. Our approach utilizes a dual-model structure: an encrypted model and a plaintext model, which are integrated through a fusion mechanism. We first identify the most privacy-sensitive features through principal component analysis (PCA) and encrypt this portion of the model, ensuring it \textit{remains encrypted throughout training} without decryption. This selective encryption strategy balances privacy and efficiency by minimizing HE overhead while preserving model security. \revised{We note here that using PCA for feature selection in this work is purely illustrative and \sys is agnostic to the choice of feature-selection method. Similarly, \sys can be integrated into any encrypted FL work that supports alternative neural network structures or utilizes various packing strategies, as its primary contribution is the fusion component.}

Unlike traditional secure aggregation, \sys prevents any party—including both clients and the server—from compromising this protected model segment. Meanwhile, the remaining portion that does not expose sensitive information is trained using the plaintext model. \sys offers benefits along two key dimensions. First, it optimizes HE operations by replacing expensive rotations with redundant parallel computations and reducing the need for bootstrapping through a fusion mechanism. Second, by training both an encrypted and a plaintext model, \sys achieves a practical balance between privacy and computational efficiency, outperforming state-of-the-art fully encrypted FL pipelines. Consequently, this also reduces bandwidth requirements and further enhances the practicality of HE-based FL for large-scale applications. 

In this paper, we demonstrate how our selective model portion can prevent reconstruction attacks through the improved deep leakage from gradients (iDLG) attack~\cite{zhao2020idlgimproveddeepleakage}. Then, our analysis indicates that the model utility is preserved and \sys scales linearly with the number of parties in the FL system. Our key contributions are as follows:

\begin{itemize}
    \item An efficient feature-selection-based selective homomorphic encryption approach that encrypts only the most privacy-sensitive features, reducing both computational and communication overhead by minimizing the volume of encrypted data transmitted.
   \item  A novel fusion design for privacy-preserving FL where efficiency gains arise naturally: reducing the encrypted feature dimension lowers HE-side overhead, while an inexpensive plaintext support network adds capacity without incurring any HE cost.
    \item A novel packing strategy for dense neural networks, applied to both end-to-end training and inference, that calculates network padding requirements from a global perspective to eliminate redundant rotations in deeper layers and support single-ciphertext mini-batching.
    
\end{itemize}

Together, by addressing privacy as a feature-level knowledge exposure problem, this work contributes to the broader goals of privacy-aware data analytics and knowledge discovery in distributed systems. 
To the best of our knowledge, \sys is the first system to perform fully encrypted model training while incorporating a fusion mechanism between encrypted and plaintext model components.



%% file: related.tex
\section{Related Work}
In this section, we review prior work on secure aggregation in FL and homomorphic encryption-based solutions.

\subsection{Secure Aggregation}

Secure aggregation has become a foundational component in FL to ensure the privacy of individual client updates. Bonawitz et al.~\cite{bonawitz2017practical} enable a server to compute the sum of client updates without learning any individual update. This is achieved by having clients mask their updates with random pads that cancel out in aggregate, effectively implementing an MPC for summation. Additive HE was also employed in BatchCrypt~\cite{zhang2020batchcrypt}, an efficient system for cross-silo FL. BatchCrypt applies Paillier encryption to randomly projected, quantized gradients, reducing communication and encryption costs with minimal accuracy loss. Park and Lim~\cite{park2022privacy} utilize a cloud server and computation provider to collaboratively aggregate encrypted local model parameters—each encrypted with different keys using a distributed HE scheme that supports partial decryption without revealing individual client data. Gronberg et al. introduced BlindFL~\cite{gronberg2025blindfl}, which segments model updates and applies FHE selectively, balancing security and computational efficiency. 
FL schemes have also explored cryptographic aggregation techniques. For instance, Lu et al.~\cite{lu2023privacy} propose a decentralized approach that uses consensus protocols and secret sharing to aggregate model updates over dynamic networks without relying on a central server. More recently, Hu and Li~\cite{10506637} propose a FL framework with selective homomorphic encryption, where clients apply HE to only a subset of the model parameters. In this sense, their approach is conceptually similar to ours in its selective use of encryption. However, a key difference is that their work focuses solely on aggregation, whereas we consider the entire training process under encryption. Overall, secure aggregation protects privacy only from the server’s perspective, as the model updates on the client side remain vulnerable to model inversion attacks. 




\subsection{Homomorphic Encryption (HE)}
HE enables a party (or parties) to compute on ciphertexts such that the decrypted result matches the outcome of computations on the original plaintexts. In machine learning, HE enables model training and inference directly on encrypted data or models, preserving data privacy throughout the process. 
Below, we review recent works that leverage HE in both centralized and FL settings, highlighting their core contributions, assumptions, and novel techniques.

\descr{Encrypted Centralized Training or Inference.} Early research on HE in machine learning focused primarily on centralized training or inference tasks~\cite{bachrach16,8260844,9160866}. A notable example is CryptoNets, which demonstrated that neural networks can perform predictions on encrypted images with reasonable accuracy and efficiency~\cite{bachrach16}. Extending HE to training is far more challenging due to the iterative and complex computations required by gradient-based learning. Nandakumar et al.~\cite{Nandakumar_2019_CVPR_Workshops} presented one of the first demonstrations of training a neural network entirely on encrypted data by implementing stochastic gradient descent (SGD) to train a simple fully-connected network. 

\descr{Encrypted Federated Learning (FL).} 
Beyond secure aggregation, which does not protect against client-side model attacks, several works explore fully encrypted training in FL, ensuring that neither the server nor clients ever see the model in plaintext during training. 
The first example for neural network training was POSEIDON~\cite{sav2021poseidon}, which employs MHE and enables under-encryption training in FL. By distributing the secret key among multiple parties, it ensures that no single entity can decrypt the model independently. This design offers robustness against collusion in the semi-honest model, remaining secure even if up to $N-1$ parties collude. However, their approach remains impractical for FL environments, requiring hours to days of training time. The complexity of training is further improved in \textsc{Hercules}~\cite{hercules}.

Unlike prior works, our framework, \sys, is the first to support \emph{selective encryption} within an end-to-end encrypted FL system. This selective approach enables \sys to encrypt only privacy-sensitive data, leaving non-sensitive information unencrypted. As a result, \sys maintains strong privacy guarantees against client-side attacks while significantly reducing computational overhead compared to systems that encrypt all data indiscriminately.

%% file: background.tex
\section{Background}\label{sec:background}
In this section, we provide the preliminaries on federated learning, principal component analysis, multiparty homomorphic encryption (MHE), and fusion mechanisms. 
We provide the frequently used notation and abbreviations in Supplementary Material A.
\subsection{Federated Learning (FL)} \label{sec:FLbackground}
FL~\cite{federatedLearning1,Konency2016fed} is a decentralized approach that trains a shared model by aggregating locally computed updates on client devices, minimizing the need to transfer large or sensitive data to a central server. A central server initializes a global model and distributes it to a randomly selected subset of participating clients. Each client then trains the model locally using its private dataset before transmitting the updated model parameters back to the server. These locally trained updates are aggregated at the server using various aggregation techniques, e.g., with FedAvg (see Algorithm~\ref{alg:fedavg}) employing weighted averaging to refine the global model. Let $W^{k, \, t}$ denote the locally updated model weights of client $k$, at iteration $t$. The aggregation process at the server is performed as follows: $
\Delta W^{t} = \sum_{k \in K} \frac{n_k}{n_t} W^{k, \, t}
$
where $n_k$ is the number of data samples held by client $k$, and $n_t = \sum_{k \in K} n_k$ represents the total number of samples across all clients at iteration $t$. FedAvg allows multiple local updates per client before transmitting model updates to the server. This iterative process continues until the model converges or a predefined number of $T$ iterations. FL thus provides key advantages, such as enhanced privacy, reduced communication costs, and improved scalability for distributed machine learning applications. 

\begin{algorithm}
\caption{Federated Averaging (FedAvg) Algorithm \cite{federatedLearning1}}
\label{alg:fedavg}
\begin{algorithmic}[1]
    \State \textbf{Server executes:}
    \State Initialize global model $W^1$
    \For{each iteration $t = 1, 2, \dots, T$} \Comment{Global iterations}
        \For{each client $k \in K$ \textbf{in parallel}} 
            \State $\nabla W^{k,t} \gets \texttt{ComputeLocalGradients}(W^{k,t})$
        \EndFor
        \State \textit{\underline{Aggregation (Server-side):}}
        \Statex\hspace{\algorithmicindent}$n_t      \gets \sum_{k\in K} n_k$
        \Statex\hspace{\algorithmicindent}$\Delta W^{t} \gets \sum_{k\in K}\frac{n_k}{n_t}\,W^{k,t}$
        \State \textit{\underline{Model Update (Client-side):}}
        \Statex\hspace{\algorithmicindent}$W^{k,t+1} \gets W^{k,t} - \eta \, \Delta W^{k,t}$
    \EndFor
    \State \textbf{Return:} Final global model $W^T$
\end{algorithmic}
\end{algorithm}



\subsection{Principal Component Analysis (PCA)} \label{sec:PCAbackground}

PCA is a traditional dimensionality reduction technique in machine learning and statistical analysis~\cite{baldiPCA1989}. It projects high-dimensional data into a lower-dimensional representation while preserving as much variance as possible. PCA is extensively used in applications such as feature extraction, data visualization, and noise filtering. The method is particularly beneficial in scenarios where features exhibit high redundancy or strong correlations among features. This can help with convergence, and  prevent overfitting.

PCA operates by identifying a set of orthogonal axes, called principal components, that maximize variance in the dataset. Given an input matrix X, PCA first standardizes the data by subtracting the mean ($\mu$) and dividing by the standard deviation ($\sigma$) $X_s = \frac{X - \mu}{\sigma}$. Then, the covariance matrix is computed. The principal components correspond to the eigenvectors of this covariance matrix (i.e. direction of variance), and their associated eigenvalues indicate the amount of variance captured (i.e. importance of the vector). The transformation is performed by projecting the original data onto the top $i$ eigenvectors, forming a lower-dimensional representation $Z = X_s W^i$ where $W^i$ is the matrix of the top $i$ eigenvectors, and Z is the projected data in the reduced space. In this work, we leverage PCA to identify the most privacy-sensitive features motivated by transforming data to a correlated space and by prior work~\cite{ratra2022big}.

\subsection{Multiparty Homomorphic Encryption} \label{sec:MHEbackground}

Multiparty Homomorphic Encryption (MHE) is a cryptographic method that extends traditional HE to a collaborative setting involving multiple parties~\cite{mouchet2021multiparty}. It allows a group of parties to perform computations on their encrypted data without revealing the underlying plaintexts to each other or to an external entity. This is particularly useful in scenarios where data privacy is paramount, such as collaborative data analysis across different organizations. In MHE, each party encrypts data using a shared public key, enabling the evaluation of functions directly over the aggregated ciphertexts. Decryption typically requires the cooperation of all or a threshold number of parties, ensuring that no single party can independently access the decrypted result ~\cite{mouchet2023multiparty}.

For MHE, we use the CKKS scheme over a cyclotomic ring $\mathcal{R} = \mathbb{Z}[X]/(X^\mathcal{N} + 1)$ where $\mathcal{N}$ is a power of 2. Each ciphertext $\mathbf{c}$ encodes a vector of complex numbers $\mathbf{z} \in \mathbb{C}^{\mathcal{N}/2}$ through an encoding function $\tau: \mathbb{C}^{\mathcal{N}/2} \rightarrow \mathcal{R}$. The underlying plaintext space (i.e. the padded representation of the ciphertext) consists of vectors in $\mathbb{C}^{\mathcal{N}/2}$, where homomorphic operations correspond to element-wise operations on these vectors. In summary, each ciphertext in CKKS provides $\mathcal{N}/2$ slots for encoding values. We use the following functionalities throughout this paper:

\begin{itemize}
    \item \textbf{Key Generation}: All parties collaboratively generate their individual secret keys, which then jointly produce the public encryption and evaluation keys.
    \item\textbf{Collective Bootstrapping}: A newly generated ciphertext starts with a level of $L$. However, each homomorphic operation reduces this level, eventually reaching a point where decryption may fail due to accumulated noise. Collective bootstrapping refreshes the ciphertext to enable further computations.
    \item \textbf{Encryption}: Using the public key, a party encrypts its plaintext data.
    \item \textbf{Evaluation}: Involves computations over the ciphertexts using the evaluation key, producing an encrypted result.
    \item \textbf{Decryption}: Parties collaboratively decrypt the resulting ciphertext to obtain the plaintext output.
\end{itemize}

In our work, we leverage MHE to facilitate privacy-preserving machine learning across multiple organizations. Each organization encrypts its data locally using the shared public key, allowing collaborative computations on the combined datasets without exposing individual records. The final model or result is decrypted only through the joint effort of authorized parties, ensuring that sensitive information remains confidential throughout the process. This approach aligns with data protection regulations and addresses concerns related to data sharing among organizations.

\subsection{Fusion Mechanisms} \label{sec:FMbackground}

Deep Model Fusion (DMF) is a machine learning technique where the parameters of multiple models are integrated, enabling more robust and accurate decision-making~\cite{gao2020survey}. 
Typical DMF consists of multiple sub-networks that process distinct input modalities or representations, followed by a fusion mechanism that combines intermediate latent representations before making a final prediction. Fusion networks are trained like standard neural networks, using iterative forward and backward passes. Each iteration consists of processing a batch of data through the fusion pipeline, and an epoch refers to a complete pass over the dataset.

DMF can take on different names based on where the fusion is applied. Early (or low-level) fusion happens when the first few layers or the raw data from different modalities are combined before feature extraction. This allows the model to learn joint feature representations directly from the input space, potentially capturing low-level correlations between modalities. Late (or high-level) fusion, on the other hand, occurs after each modality has undergone independent feature extraction. The outputs of different sub-networks are concatenated or aggregated at the decision (score) level, allowing the model to leverage high-level representations from each modality before making a final prediction. Intermediate fusion (or mid-level fusion) represents a hybrid approach where feature representations from different modalities are combined at one or more points within the network, typically after initial feature extraction but before the final decision layer. In this study, we adopt score-level fusion, combining the logits from individual sub-networks via weighted averaging just before the final prediction stage.


%% file: method.tex
\section{Proposed Method}
Here, we provide the details of our proposed framework. First, we define the system and threat model. Next, we outline our system overview and feature selection process. Finally, we explain our model training, including HE operations, in detail. 

\subsection{System and Threat Model}
\textbf{System Model.} We consider an FL setting in which N parties collaboratively train a neural network model without sharing their local datasets. Once training is complete, the final model is hosted either by one of the N parties or an external entity, such as a cloud service provider. Throughout the process, all parties are responsible for safeguarding data privacy and preventing reconstruction attacks at every stage—communication, training, and prediction—where potential attackers could include an individual party, the server, or a collusion of multiple parties.

\par\noindent\textbf{Threat Model.} We operate under an honest-but-curious setting with $K$ parties, where our setup assumes the possibility of collusions involving up to  $K-1$ parties, all of whom adhere to the protocol but may attempt to infer information from other parties' data. Our primary goal is to prevent reconstruction and inference attacks during FL training, where the server or clients might exploit the model or gradients to infer information about the remaining parties' data. However, attacks that exploit prediction outputs are considered out of the scope of this work, as our focus is not on the prediction phase.
\subsection{System Overview} \label{sec:systemoverview}
In \sys, we leverage PCA to identify and select the most privacy-sensitive features in the dataset. This approach is motivated by the fact that many raw data features exhibit high correlations, which can be exploited by attackers for reconstruction. Recent research has also shown that PCA can serve as an effective method for mitigating privacy threats~\cite{ratra2022big}. \revised{While we use PCA in this work, it serves only as a demonstrative feature-selection mechanism. The feature-selection stage in HADES is fully modular, and any alternative algorithm can be plugged in without changing the rest of the system.}

\begin{figure*}[t] 
    \centering
\includegraphics[width=\textwidth,height=\textheight,keepaspectratio]{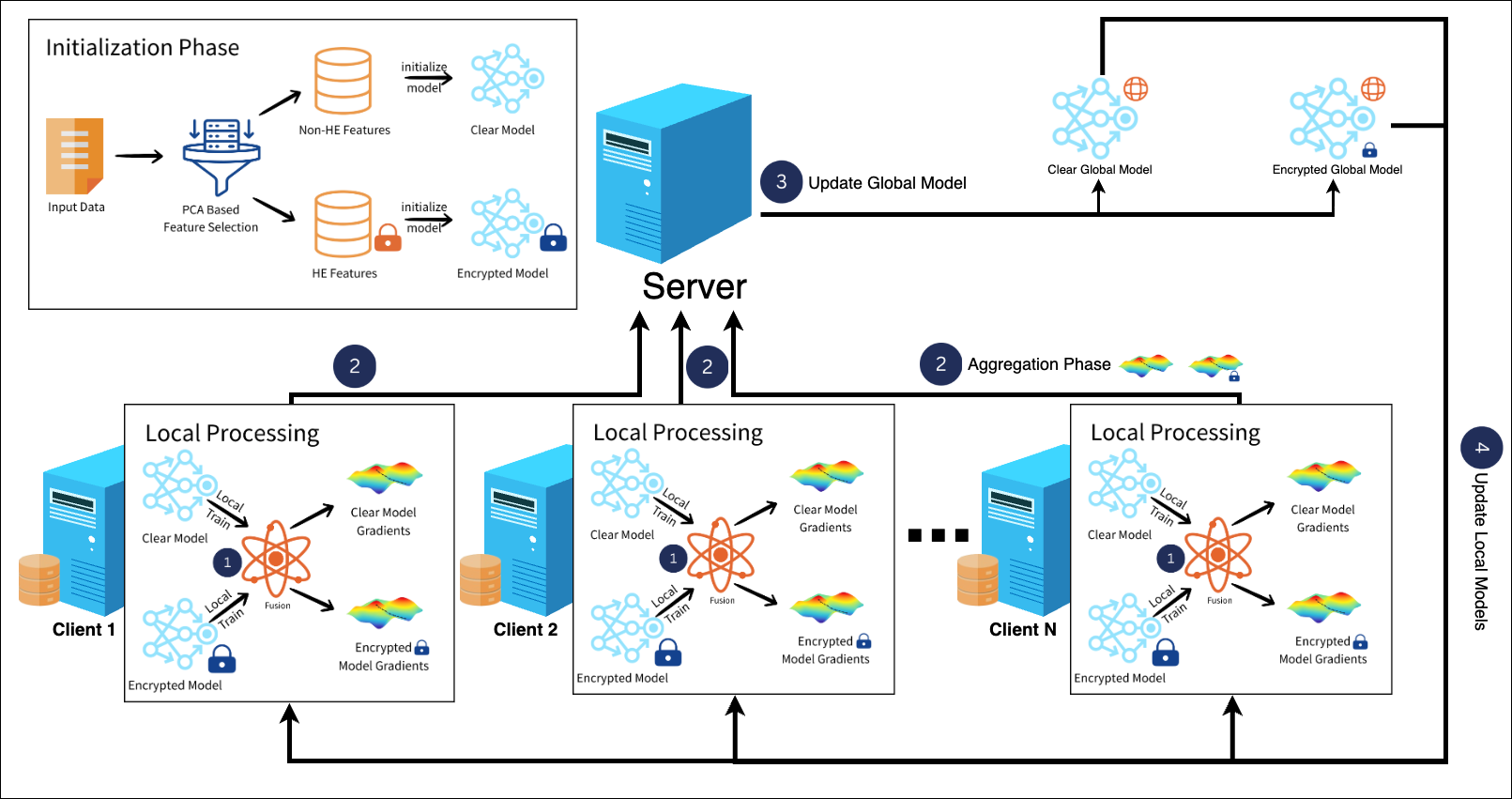}
    \caption{\sys's system overview. The initialization phase involves selecting the clear-text and encrypted (privacy-sensitive) components of the model using PCA-based feature selection. Overall, \sys's online training consists of local computations in an FL setting, where two models—one operating on clear-text data and the other on encrypted data are trained locally. Their updates are then aggregated on the server and subsequently fused into a single model (via fusion) during local post-processing. \revised{The encrypted network is optimized using the fused loss while keeping all HE parameters and logits encrypted, while the plaintext network is trained with its own plaintext loss, without requiring any decryption or access to the encrypted parameters.}}
    \label{fig:system}
\end{figure*}

The selected high-sensitivity features are processed within an encrypted environment, ensuring their confidentiality during computation. In contrast, the remaining, less-sensitive features are handled in plaintext, optimizing computational efficiency. To maintain both security and performance, we employ a fusion strategy that seamlessly integrates logits from both processing streams. \revised{These fused logits guide the learning on the encrypted branch while preserving end-to-end encryption of its weights. Separately, the plaintext branch uses its own loss, calculated from only plaintext logits, and requires no decryption of the HE parameters.} Figure~\ref{fig:system} provides an overview of \sys's high-level workflow.
We also summarize \sys's federated training procedure in Algorithm~\ref{algo:hades}, which begins with an initialization phase for feature selection for encryption (Lines 1–2). We detail this procedure in the next section (Section~\ref{sec:featureSelection}). Then, clients perform local training (Lines 5-11) followed by server-side aggregation (Lines 12-14), which are detailed in Section~\ref{model-training}.

\subsection{Feature Selection for Encryption}\label{sec:featureSelection}
Our proposed FL system, \sys, begins with feature selection (Algorithm~\ref{algo:hades}, Lines 1–2). Let $\mathcal{F} = \{ f_1, \ldots, f_d \}$ be the original feature set. We split the feature set into $\mathcal{F}_{HE}$, the subset whose parameters are handled by the HE encrypted model, and $\mathcal{F}_{P}$, the subset processed in plaintext. Consequently, $|\mathcal{F}_{HE}|$ and $|\mathcal{F}_{P}|$ denote the \textit{number} of encrypted and plaintext features, respectively, throughout this paper.

 PCA is used to find principal components of length $|\mathcal{F}|$ that describe most of the variance. From $\mathcal{F}$, we select $\mathcal{F}_{HE}$ to be the top-$i$ principal components. The selection of $i$ is determined based on three factors: the proportion of variance explained, network configuration to accommodate ciphertext size, and the desired utility of the resulting model. As demonstrated through extensive experiments in Section~\ref{sec:q1}, our approach effectively mitigates reconstruction attacks, depending on the number of principal components selected.

The ratio
$
\frac{\lvert \mathcal{F}_{HE}\rvert}{\lvert \mathcal{F}_{P}\rvert}
$
depends on the dataset and task requirements: increasing $\lvert \mathcal{F}_{HE}\rvert$ offers stronger privacy guarantees but incurs heavier computational overhead. Furthermore, the ring size $\mathcal{N}$ limits the amount of data that can be protected, as only $2^{\mathcal{N}-1}$ elements fit per ciphertext slots (see Section~\ref{sec:MHEbackground}). Both the encrypted features and the model weights must stay within that capacity. This constraint, in turn, forces the layer sizes and intermediate operations to be chosen so every tensor remains representable in ciphertext space. As an example, a matrix multiplication of dimensions $N \times M$ and $M \times K$ produces an $N \times K$ output, but this operation requires $N \times M \times K$ intermediate elements in HE multiplication. For homomorphic operations to remain valid, the condition
$
N \times M \times K \le 2^{\mathcal{N}-1}
$
must hold, reflecting the ciphertext's capacity limit. As a simple example, for batch size $B$ on a single-layer network with a single output neuron, the effective capacity becomes
\[
\lvert \mathcal{F}_{HE}\rvert \le  \frac{2^{\mathcal{N}-1}}{B}.
\]


To extract information about the training data, an adversary would need to approximate the inverse PCA transformation, which remains encrypted. Even if decryption were possible, (which is prevented in \sys through MHE), the adversary would still require the transformation parameters to reconstruct the original data, adding an extra layer of security.
While this increases the complexity of an attack, our model does not rely solely on the inaccessibility of the PCA transformation, but primarily on the security guarantees of MHE encryption.

Notably, PCA can also be used to guide feature selection without requiring the model to be trained using principal components as features. In this scenario, the absolute contributions of the top-k principal components are aggregated to rank the original features based on their relevance. This allows for informed feature selection while preserving the original feature space. However, we choose to use PCA-derived features, as they do not necessarily lead to performance degradation during training (see Section~\ref{sec:q1}). 

Finally, we note that privacy-preserving federated PCA computation is beyond the scope of this paper, as it has been well-addressed in the existing literature. Based on the literature for implementing PCA in a federated setting, we propose two possible approaches. First, each party can independently apply PCA on its local dataset to determine its private feature set, then leverage Private Set Intersection to establish a consensus on the selected features. Alternatively, clients can leverage privacy-preserving multi-party PCA solutions, such as PPPCA \cite{PPPCA} or SF-PCA \cite{SP-PCA}, to securely compute principal components in a collaborative manner while maintaining privacy.

\begin{algorithm}[t]
\small
\caption{HADES Training Algorithm.}
\label{algo:hades}
\begin{algorithmic}[1]
\Statex \textbf{Initialization:}
\State Parties agree on a secure PCA transformation.
\State Each client $k$ applies the PCA transformation, splitting data into $X_{HE}$ and $X_{P}$, with feature sets $\mathcal{F}_{HE}$ and $\mathcal{F}_{P}$ respectively.
\State Initialize local models $W_{HE}^k$ and $W_{P}^k$.

\Statex \textbf{Training Loop:}
\For{Each iteration $t$}
    \Statex \textit{\underline{Local Training (Client-side):}}
    \For{Each client $k$ in $\{1, \dots, K\}$}
        \State Forward pass of $X_{P}$ through $W_{P}^{k, t}$, output logits $\mathbf z_{P}^{k,t}$.
        \State Forward pass of $X_{HE}$ through $W_{HE}^{k, t}$, output \revised{encrypted} logits $\mathbf z_{HE}^{k,t}$.
        \revised{
        \State Compute fused logit $\bar{\mathbf z}_{HE}^{k,t} = \alpha\,\mathbf z_{HE}^{k,t} + (1-\alpha)\,\mathrm{HE}(\mathbf z_{P}^{k,t})$. Set plaintext logit $\bar{\mathbf z}_{P}^{k,t} = \mathbf z_{P}^{k,t}$.
        \State Compute objectives 
        $\mathcal{L}_{HE}^{k,t} = \ell(\bar{\mathbf z}_{HE}^{k,t}, \mathbf y^{k})$
        and
        $\mathcal{L}_{P}^{k,t} = \ell(\bar{\mathbf z}_{P}^{k,t}, \mathbf y^{k})$.
        \State Calculate gradients $\nabla W_{HE}^{k,t}$ w.r.t. $\mathcal{L}_{HE}^{k,t}$ and $\nabla W_{P}^{k,t}$ w.r.t. $\mathcal{L}_{P}^{k,t}$.
        }
        \State Send $\bigl[\nabla W_{HE}^{k,t}, \nabla W_{P}^{k,t}\bigr]$ to server.
    \EndFor
    
    \Statex \textit{\underline{Aggregation (Server-side):}}
    \State Aggregate encrypted updates: $\Delta W_{HE}^{t} \gets \sum_{k=1}^K \nabla W_{HE}^{k,t}$.
    \State Aggregate plaintext updates: $\Delta W_{P}^{t} \gets \sum_{k=1}^K \nabla W_{P}^{k,t}$.
    \State Send aggregated $\Delta W_{HE}^{t}, \Delta W_{P}^{t}$ back to clients.
    
    \Statex \textit{\underline{Model Update (Client-side):}}
    \For{Each client $k$ in $\{1, \dots, K\}$}
        \State $W_{HE}^{k,t+1} \gets W_{HE}^{k,t}-\eta \, \Delta W_{HE}^{t}$
        \State $W_{P}^{k,t+1} \gets W_{P}^{k,t}-\eta \, \Delta W_{P}^{t}$.
    \EndFor
\EndFor
\end{algorithmic}
\end{algorithm}

\subsection{Model Training}
\label{model-training}

We propose a fusion-based approach that divides model parameters into homomorphically encrypted ($W_{HE}$) and plaintext ($W_{P}$) sub-networks. After agreeing on the selected feature set for encryption $\mathcal{F}_{HE}$ (see Section~\ref{sec:featureSelection}), each client initializes local models $W_{HE}^k$ and $W_{P}^k$ with input dimensions $|\mathcal{F}_{HE}|$ and $|\mathcal{F}_{P}|$ respectively (Algorithm~\ref{algo:hades} Line 3). 

Each client $k$ applies forward and backward passes on their local data on both $W_{HE}^k$ and $W_{P}^k$ (Lines 6-7). \revised{Weights and g}radients for $W_{HE}^k$ are never decrypted during forward and backward passes. For \revised{the} HE operations, we require efficient packing and multiplication strategies to replace their vanilla counterparts in plaintext training. Details on how we define and use HE operations in forward and backward passes, packing, and bootstrapping are presented in Supplementary Material B.

After the forward pass, each network generates partial predictions based on its respective feature set. \revised{From the encrypted and plaintext logits, we form two objectives: (i) an encrypted and fused objective used to train the encrypted sub-network, and (ii) an independent plaintext objective used to train the plaintext sub-network. 
Specifically, let $z_{HE}(\mathbf{x}_i;W_{HE})$ and $z_{P}(\mathbf{x}_i;W_{P})$ denote the encrypted and plaintext logits, respectively. 
We define the fused HE logit as a weighted ciphertext--plaintext combination (without decrypting $z_{HE}$), while the plaintext logit is computed solely from $z_P$:
\begin{align}
\bar{z}_{HE}
&= \alpha\, z_{HE} + (1-\alpha)\,\mathrm{HE}(z_{P}) \\
\bar{z}_{P}
&= z_{P}.
\end{align}

We use the residual objective $\ell(\hat{z},y) := \hat{z}-y$, yielding
\begin{align}
\mathcal{L}_{HE} &= \ell\!\bigl(\bar{z}_{HE}, y_i\bigr)
= \bar{z}_{HE}(\mathbf{x}_i) - \mathrm{HE}(y_i), \\
\mathcal{L}_{P}  &= \ell\!\bigl(\bar{z}_{P}, y_i\bigr)
= \bar{z}_{P}(\mathbf{x}_i) - y_i.
\end{align}
where $\alpha\in[0,1]$ controls the relative contribution of each sub-network to the fused objective (Line 8).} While we do not explore this within the scope of this study, \emph{intermediate fusion} can be used for additional flexibility: the plaintext hidden features $h_{P}$ are encrypted on the fly and then concatenated (or used in an attention mechanism) with $h_{HE}$.

Notably, at no point are the homomorphic parameters decrypted. Both the parameters and their gradients remain encrypted throughout the entire training process. The only unencrypted components of the network are the intentionally unencrypted model parameters. \revised{Both encrypted and plaintext model losses are} retained locally by each client and are never shared with the server.

We use the standard communication protocol used in FL, where, in iteration $t$, client $k$ computes local updates (Line 10) $
\nabla W_{HE}^{k,t}, 
\,
\nabla W_{P}^{k,t},
$ 
and sends them to the server (Line 11). The server aggregates these two parameters separately (Lines 13-14), then broadcasts the updates back to clients (Line 15). This structure ensures that the encrypted model part and gradients are always secure during local training, communication, and server aggregation. Each client then applies the aggregated model parameters and proceeds to the next round (Lines 16-19). Below, we detail the ciphertext operations required to enable training on the HE-encrypted network $W_{HE}$.

\subsubsection{Approximation of activation functions}

In HE, non-polynomial activation functions such as the sigmoid cannot be directly evaluated on encrypted data. To address this, we approximate them \revised{with low-degree polynomials obtained via least-squares fitting over a chosen input interval. Concretely, we sample the target function at a set of points within the approximation range and solve for polynomial coefficients that minimize the squared error between the polynomial and the target values. At runtime, the activation is evaluated on the encrypted layer output using the pre-computed coefficients via standard polynomial evaluation.} During the backward pass, we use the derivative of the fitted polynomial. Alternative approximation schemes can be seamlessly integrated into \sys; however, the choice and evaluation of approximation methods is beyond the scope of this work.



%% file: Tables/pca_variance_table.tex
\begin{table*}[t]
\centering
\small
\caption{Cumulative variance explained (\%) by PCA across datasets at different feature selection thresholds, alongside test accuracy (\%) for \textbf{Baseline} non-fusion network with only encrypted features and \textbf{\sys}. Missing values (-) indicate cases where the feature count exceeds the maximum available features for that dataset.}
\label{tab:pca_baseline_comparison}
\resizebox{\textwidth}{!}{
\begin{tabular}{c|ccc|ccc|ccc}
\toprule
Feature Count
& \multicolumn{3}{c|}{BCD}
& \multicolumn{3}{c|}{MNIST}
& \multicolumn{3}{c}{SVHN} \\
\cmidrule(lr){2-4}\cmidrule(lr){5-7}\cmidrule(lr){8-10}
& Var. (\%) & Baseline Acc. (\%) & \sys Acc. (\%)
& Var. (\%) & Baseline Acc. (\%) & \sys Acc. (\%)
& Var. (\%) & Baseline Acc. (\%) & \sys Acc. (\%) \\
\midrule
1    & 98.14  & 94.74 & 93.57  & 9.76  & 20.68 & 90.44  & 57.91 & 19.59 & 66.06 \\
2    & 99.79  & 94.74 & 92.98  & 16.92 & 41.93 & 91.14  & 63.62 & 19.59 & 66.68 \\
4    & 99.99  & 95.91 & 97.08  & 28.48 & 60.70 & 89.56  & 72.96 & 24.89 & 60.78 \\
8    & 100.00 & 98.25 & 97.66  & 43.87 & 84.15 & 90.67  & 79.95 & 30.97 & 54.15 \\
16   & 100.00 & 95.32 & 95.32  & 59.58 & 91.39 & 92.64  & 86.41 & 57.00 & 59.75 \\
32   & -      & -     & -      & 74.49 & 93.44 & 93.82  & 91.60 & 67.23 & 69.23 \\
64   & -      & -     & -      & 86.28 & 94.60 & 94.54  & 96.00 & 70.61 & 72.01 \\
128  & -      & -     & -      & 93.68 & 94.52 & 94.64  & 98.45 & 70.66 & 72.16 \\
256  & -      & -     & -      & 97.94 & 94.43 & 94.73  & 99.47 & 70.25 & 71.45 \\
512  & -      & -     & -      & 99.94 & 94.46 & 94.67  & 99.86 & 70.44 & 71.40 \\
1024 & -      & -     & -      & -     & -     & -      & 99.98 & 69.91 & 71.67 \\
2048 & -      & -     & -      & -     & -     & -      & 100.00 & 69.70 & 72.12 \\
\bottomrule
\end{tabular}
}
\end{table*}

%% file: Tables/recon_attack_table.tex
\begin{table}[htbp]
\centering
\caption{Reconstruction Quality Metrics (RMSE, PSNR, SSIM, LPIPS) by Dataset and $|\mathcal{F}_{HE}|$. Each cell shows values for the random-initialization attack. $|\mathcal{F}_{HE}|$ controls the number of features selected via PCA to be encrypted. $|\mathcal{F}_{HE}| = 0$ is the perfect reconstruction case where \sys is not used. $|\mathcal{F}_{HE}| = |\mathcal{F}| - 1$ denotes the case where only a single feature can be used for reconstruction, which sets the worst case values of the metrics for a given dataset.}
\resizebox{\columnwidth}{!}{
\begin{tabular}{p{3.4cm}|c|cccccc}
\toprule
\textbf{Dataset} & \textbf{Metric} & 
\multicolumn{6}{c}{$|\mathcal{F}_{HE}|$} \\
\cmidrule(lr){3-8}
& & \textbf{0} & \textbf{16} & \textbf{64} & \textbf{256} & \textbf{1024} & \textbf{$|F| - 1$} \\
\midrule
\multirow{4}{*}{\parbox{3.4cm}{BCD}} & RMSE & 0.00 & 92.99 & -- & -- & -- & 98.86 \\
 & PSNR (dB) & $\infty$ & 13.51 & -- & -- & -- & 8.60 \\
 & SSIM & 1.00 & 0.86 & -- & -- & -- & 0.65 \\
 & LPIPS & N/A & N/A & -- & -- & -- & N/A \\
\midrule
\multirow{4}{*}{\parbox{3.4cm}{MNIST}} & RMSE & 0.00 & 0.22 & 0.29 & 0.42 & -- & 0.63 \\
 & PSNR (dB) & $\infty$ & 13.48 & 11.09 & 8.92 & -- & 5.97 \\
 & SSIM & 1.00 & 0.48 & 0.26 & 0.11 & -- & 0.04 \\
 & LPIPS & 0.00 & 0.17 & 0.29 & 0.49 & -- & 0.64 \\
\midrule
\multirow{4}{*}{\parbox{3.4cm}{SVHN}} & RMSE & 0.00 & 0.17 & 0.20 & 0.26 & 0.39 & 0.62 \\
 & PSNR (dB) & $\infty$ & 16.21 & 14.29 & 11.96 & 8.97 & 7.31 \\
 & SSIM & 1.00 & 0.53 & 0.22 & 0.04 & 0.01 & 0.01 \\
 & LPIPS & 0.00 & 0.16 & 0.26 & 0.53 & 0.83 & 0.87 \\
\bottomrule
\end{tabular}
}

\label{tab:recon_stats}
\end{table}

%% file: Figures/recon_images.tex
\begin{figure}[htbp]
\centering
\subfloat[MNIST]{
  \includegraphics[width=0.2\linewidth]{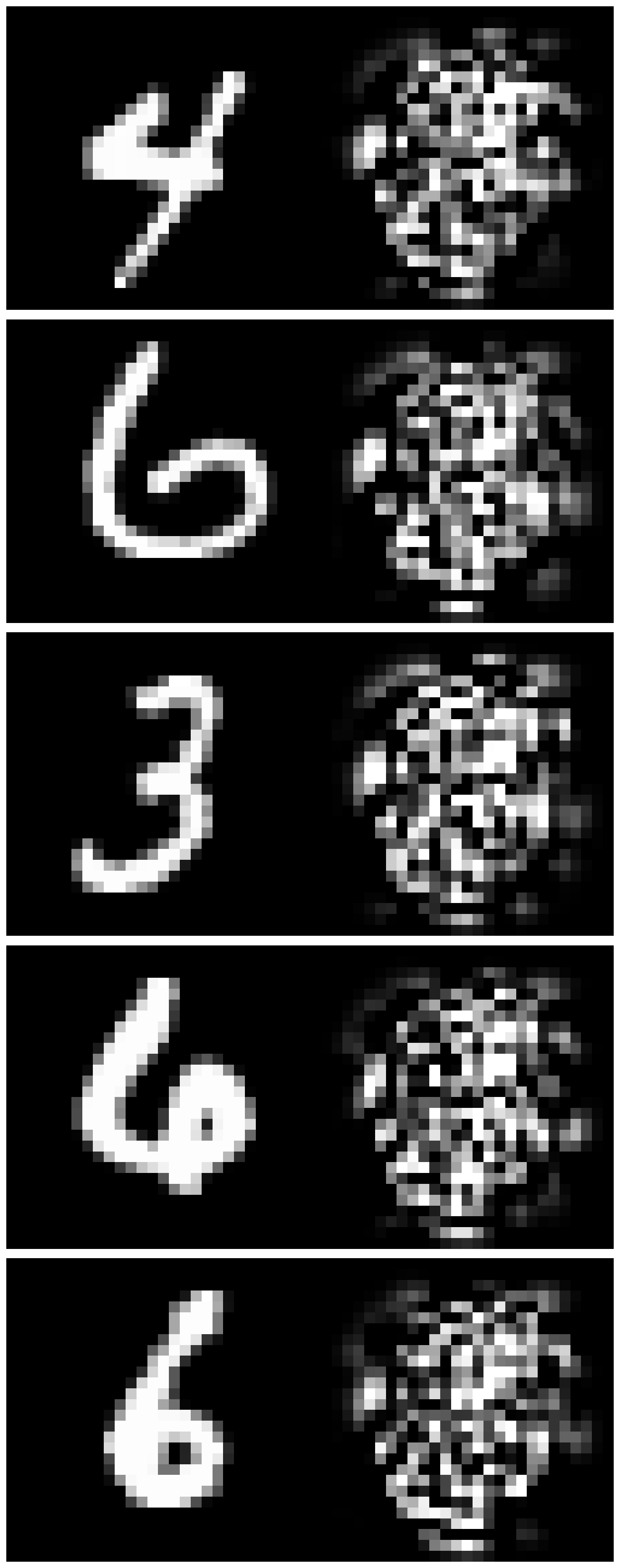}
  \label{fig:recon_mnist_fusion256}
}
\subfloat[SVHN]{
  \includegraphics[width=0.2\linewidth]{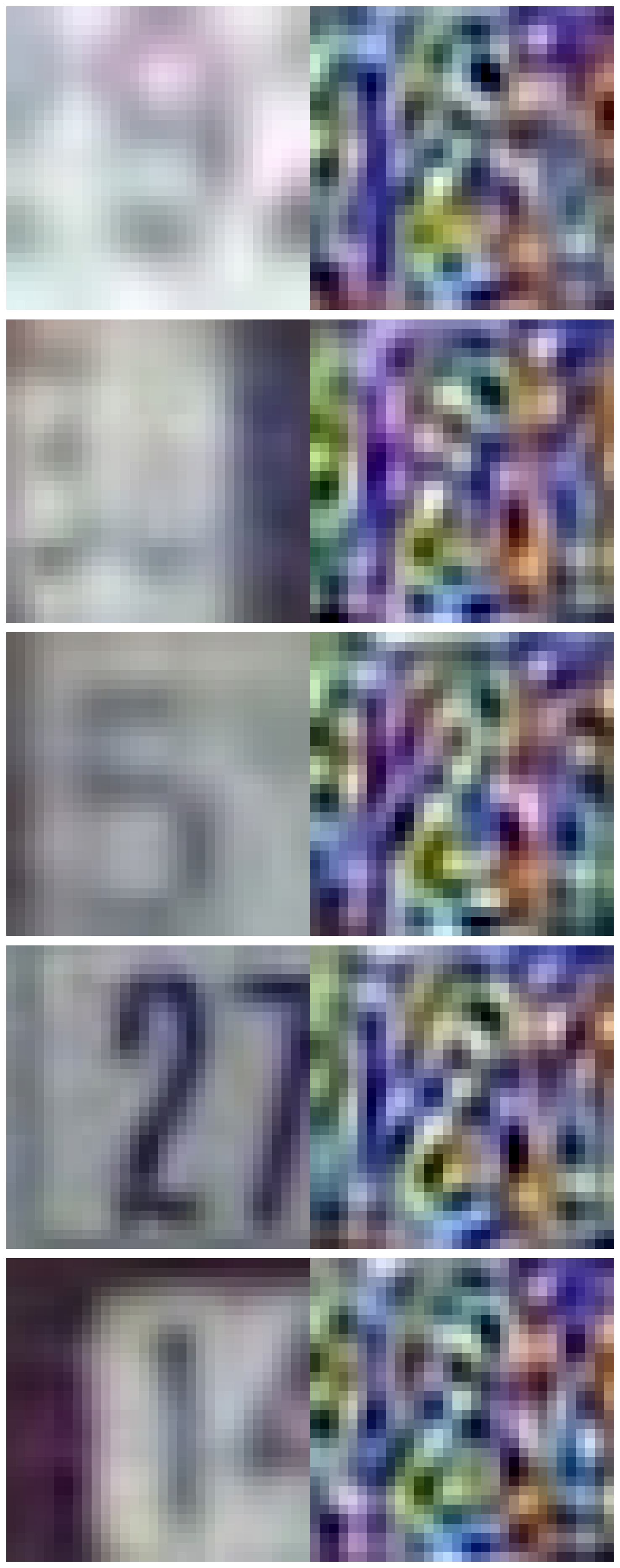}
  \label{fig:recon_svhn_fusion256}
}
\caption{Qualitative iDLG reconstructions for the fusion model with $|\mathcal{F}_{HE}|=256$ on MNIST (a) and SVHN (b).}
\label{fig:recon_fusion_256}
\end{figure}

%% file: Tables/exp2_table.tex
\begin{table}[ht!]
\caption{Test accuracies (\%) for centralized (CE) and federated (FL) training across different configurations of $\mathcal{F}_{HE}$. $\mathcal{F}_{HE} = 0$ means all data is processed with a fully plaintext model. Bold $\boldsymbol{\mathcal{F}_{HE}}$ denotes that a model portion is trained under encryption, and we set $|\mathcal{F}_{HE}|=[16, 256, 256]$ for BCD, MNIST, SVHN respectively.}
\label{tab:exp2_results}
\small
\resizebox{\columnwidth}{!}{%
\begin{tabular}{lcccc}
\toprule
Dataset & CE- $\mathcal{F}_{HE}=0$ & CE- $\mathcal{F}_{HE}$ & FL- $\mathcal{F}_{HE}$ & FL- $\mathcal{F}_{HE}$ approx. (\sys) \\
\midrule
BCD & 94.71 & 95.88 & 96.49 & 97.08  \\
MNIST & 95.00 & 94.80 & 94.82 & 94.99 \\
SVHN & 63.1 & 66.2 & 67.3 & 70.4 \\
\bottomrule
\end{tabular}
}
\end{table}

%% file: Tables/gi_per_dataset.tex
\begin{table}[ht]
\centering
\small
\caption{One-GI and Total Training timing analysis for \textsc{BCD}, \textsc{MNIST}, and \textsc{SVHN} under different batch sizes and encrypted feature set sizes ($|\mathcal{F}_{HE}|$). \emph{GI-Load} denotes the number of global iterations (GI) required to process a fixed amount of data (baseline~$X$). \emph{One-GI} is the wall-clock time for a single global iteration, and \emph{Total Training} is the per-client end-to-end training time. For comparability, we run configurations with GI-Load $=16X$ for 100 iterations and scale the iteration count linearly for smaller loads as $I_{kX}=100\cdot k/16$ (e.g., $X=100/16$). For \textsc{BCD}, we use one hidden layer with 64 units; for \textsc{MNIST} and \textsc{SVHN}, we use no hidden layers. $|\mathcal{F}|$ is 30, 784 and 3072 for \textsc{BCD}, \textsc{MNIST} and \textsc{SVHN}, respectively.}
\resizebox{\columnwidth}{!}{%
\begin{tabular}{l c c c c c}
\toprule
Dataset & Batch Size & $|\mathcal{F}_{HE}|$ & GI-Load & One-GI (s) & Total Training (s)\\
\midrule
\multirow{8}{*}{BCD} & 1  & 30 & 16X & 0.519 & 202.5\\
 & 2  & 30 & 8X  & 0.593 & 115.6\\
 & 4  & 16 & 4X  & 0.547 & 53.4\\
 & 4  & 8  & 4X  & 0.530 & 51.7\\
 & 8  & 8  & 2X  & 0.534 & 26.0\\
 & 8  & 4  & 2X  & 0.502 & 24.5\\
 & 16 & 4  & X   & 0.543 & 13.2\\
 & 16 & 2  & X   & 0.511 & 12.5\\
\hline
\multirow{6}{*}{MNIST} & 1 & 256 & 16X & 0.313 & 15337.5\\
 & 1 & 128 & 16X & 0.262 & 12816.7\\
 & 1 & 64  & 16X & 0.240 & 11738.9\\
 & 2 & 128 & 8X  & 0.301 & 7381.4\\
 & 2 & 64  & 8X  & 0.267 & 6543.6\\
 & 4 & 64  & 4X  & 0.294 & 3602.8\\
\hline
\multirow{4}{*}{SVHN} & 1 & 256 & 16X & 0.305 & 22332.2\\
 & 1 & 128 & 16X & 0.264 & 19325.8\\
 & 2 & 128 & 8X  & 0.297 & 10874.4\\
 & 4 & 64  & 4X  & 0.277 & 5071.9\\
\bottomrule
\end{tabular}%
}
\label{tab:gi_dataset}
\end{table}

%% file: supplementary.tex
\newtheorem{definition}{Definition}
\section*{SUPPLEMENTARY MATERIAL}

\subsection{Glossary}
We provide the frequently used notations and abbreviations throughout the manuscript in Table~\ref{Glossary}.

\input{glossary}

\subsection{Details of HADES's training and packing strategies}\label{sec:detailed}

\subsubsection{Matrix Representation via Packing and Padding Strategies} \label{sec:local_packing}

One key challenge in implementing HE-based neural networks is efficiently encoding (packing) matrix representations and operations while adhering to constraints imposed by HE. In this section, we provide the mathematical details of HE operations that enable neural network training under encryption. These operations encompass our packing strategies for weight matrices, as well as ciphertext masking, ciphertext rotation, and the approximation of the activation functions. 

We adopt a packing strategy similar to that introduced in~\cite{sav2021poseidon}, which alternates matrix packing based on the layer index (the layer’s position counted from the input, starting from 1), as detailed below. This strategy adapts based on the operational requirements, such as matrix multiplication and network architecture. For a given matrix $\mathbf{M} \in \mathbb{R}^{m \times n}$, we define a packing function $\phi: \mathbb{R}^{m \times n} \rightarrow \mathbb{C}^{N/2}$ that maps the matrix to a vector suitable for encryption. This function fundamentally involves flattening the matrix into a one-dimensional array and strategically inserting padding (zeros) between matrix elements or to adjust dimensions.

The flattening step can be performed in two ways: row-major or column-major ordering. Let $\pi_{\text{row}}$ and $\pi_{\text{col}}$ denote these linearization functions.

For example, consider the following matrices:
\[
\mathbf{A} = 
\begin{bmatrix}
a_{11} & a_{12}\\[6pt]
a_{21} & a_{22}
\end{bmatrix}
\xrightarrow{\pi_{\text{row}}(\mathbf{A})}
[a_{11}, a_{12}, a_{21}, a_{22}]
\]

\[
\mathbf{B} = 
\begin{bmatrix}
b_{11} & b_{12}\\[6pt]
b_{21} & b_{22}
\end{bmatrix}
\xrightarrow{\pi_{\text{col}}(\mathbf{B})}
[b_{11}, b_{21}, b_{12}, b_{22}]
\]

Their product is given by $
A \boxdot B = [\,a_{11}b_{11} + a_{12}b_{21},\;a_{11}b_{12} + a_{12}b_{22},\;a_{21}b_{11} + a_{22}b_{21},\;a_{21}b_{12} + a_{22}b_{22}\,]
$
where $\boxdot$ denotes the Hadamard (element-wise) product. The output represents a flattened form of the actual matrix product:
\[
\mathbf{A \cdot B} =
\begin{bmatrix}
a_{11}b_{11} + a_{12}b_{21} & a_{11}b_{12} + a_{12}b_{22}\\[6pt]
a_{21}b_{11} + a_{22}b_{21} & a_{21}b_{12} + a_{22}b_{22}
\end{bmatrix}
\]

Note that alternating row-major and column-major representations for matrices enables perfect alignment between the row elements of $\mathbf{A}$ and the column elements of $\mathbf{B}$. However, simply taking the Hadamard product of the resulting flattened vectors does not always yield the correct matrix product. While alternating matrix representations work in simple cases, they prove insufficient when handling consecutive matrix operations where matrix sizes are unequal or when at least one dimension in the network is not a power of 2. To address this, a crucial aspect of our packing approach involves padding matrices with zeros, extending their non-common dimensions to the smallest power of 2 that is at least as large as the largest dimension among the two matrices. This ensures that the inner dimensions of matrices in a multiplication $A \cdot B$ are conformant after any preparatory replications or expansions. It provides "empty slots" in the ciphertext, crucial for preventing data overwriting during ciphertext rotations. 

Consider the multiplication of two matrices $\mathbf{A} \in \mathbb{R}^{m \times k}$ and $\mathbf{B} \in \mathbb{R}^{k \times n}$, where $ m < n $ and $ n $ is not a power of 2. We first define the padded dimension for the non-common dimension as $\hat{n} = 2^{\lceil \log_2 n \rceil}$. Then, we compute the multiplication area $\mu$ as $m \times k \times \hat{n}$. This is the total number of ciphertext slots required to make this particular matrix multiplication. The $m \times k$ matrix $\mathbf{A}$ is padded along its $k$-dimension to become an $m \times \hat{k}$ matrix $\mathbf{\tilde{A}}$, and the $k \times n$ matrix $\mathbf{B}$ is padded along its $n$-dimension to become a $k \times \hat{n}$ matrix $\mathbf{\tilde{B}}$. Calculation of $\hat{k}=\mu/m$ for the entire network is given in Section~\ref{sec:novel_global_pad}, and it depends on the parameters of the entire network.

We formalize these transformations as follows:
\[
\xi: \mathbb{R}^{m \times k} \rightarrow \mathbb{R}^{m \times \hat{k}}, \quad \xi(A) = \tilde{A},
\]
\[
\psi: \mathbb{R}^{k \times n} \rightarrow \mathbb{R}^{k \times \hat{n}}, \quad \psi(B) = \tilde{B}.
\]
Here, $\tilde{A}$ and $\tilde{B}$ denote the padded matrices. After padding, flattened matrix representation for $\mathbf{A}$ is rotated $\lfloor \log_2 \hat{n} \rfloor$ times by $m$ elements at a time, while representation for $\mathbf{b}$ is rotated $\lfloor \log_2 m \rfloor$ times by $\hat{k}$ elements at a time. This allows for perfect local multiplication alignment while optimizing global alignment. As another example, consider these matrices:
\[
\mathbf{A} =
\begin{bmatrix}
a_{11} & a_{12}\\[6pt]
a_{21} & a_{22}
\end{bmatrix}
\quad\text{and}\quad
\mathbf{B} =
\begin{bmatrix}
b_{11} & b_{12} & b_{13}\\[6pt]
b_{21} & b_{22} & b_{23}
\end{bmatrix}.
\]

Since $m < n$ and $n$ is not a power of 2, we pad the matrices to dimensions $2 \times 8$ and $2 \times 4$, respectively. We first define $\hat{n} = 2^{\lceil \log_2 3 \rceil} = 4$, and the padded matrices are:
\begin{align*}
\tilde{\mathbf{A}}
&= \begin{bmatrix}
a_{11} & a_{12} & 0 & 0 & 0 & 0 & 0 & 0 \\[6pt]
a_{21} & a_{22} & 0 & 0 & 0 & 0 & 0 & 0
\end{bmatrix}
&&\text{(padded to $2\times8$)}\\[1em]
\tilde{\mathbf{B}}
&= \begin{bmatrix}
b_{11} & b_{12} & b_{13} & 0\\[6pt]
b_{21} & b_{22} & b_{23} & 0
\end{bmatrix}
&&\text{(padded to $2\times4$)}
\end{align*}

\begin{flushleft}
Next, we apply the row and column flattening operations:
\begin{align*}
\pi_{\text{row}}(\tilde{\mathbf{A}}) &= [\,a_{11},\, a_{12},\, 0,\, 0, 0,\, 0,\, 0,\, 0,\, \\
&\quad a_{11},\, a_{12},\, 0,\, 0,\, 0,\, 0,\, 0,\, 0] \\[6pt]
\pi_{\text{col}}(\tilde{\mathbf{B}}) &= [\,b_{11},\, b_{21},\, b_{12},\, b_{22}, b_{13},\, b_{23},\, 0,\, 0].
\end{align*}

and the rotation operations:
\begin{align*}
RR\bigl(\pi_{\text{col}}(\tilde{\mathbf{B}}),8,2\bigr) &= [\,b_{11},\, b_{21},\, b_{12},\, b_{22}, b_{13},\, b_{23},\, 0,\, 0,\, \\
&\quad b_{11},\, b_{21}, b_{12},\, b_{22},\, b_{13},\, b_{23},\, 0,\, 0].
\end{align*}

The flat multiplication $\tilde{A} \boxdot \tilde{B}$ then yields:
\begin{align*}
\tilde{A} \boxdot \tilde{B}
= [\,a_{11}b_{11} + a_{12}b_{21},\; a_{11}b_{12} + a_{12}b_{22},\; a_{11}b_{13} + a_{12}b_{23},\;
0,\; \\
a_{21}b_{11} + a_{22}b_{21},\; a_{21}b_{12} + a_{22}b_{22},\; a_{21}b_{13} + a_{22}b_{23},\; 0\,].
\end{align*}
\end{flushleft}

\noindent This flattened result corresponds to the actual padded product:
\[
\tilde{\mathbf{A}} \cdot \tilde{\mathbf{B}} =
{\small
\begin{bmatrix}
a_{11}b_{11}+a_{12}b_{21} & a_{11}b_{12}+a_{12}b_{22} & a_{11}b_{13}+a_{12}b_{23} & 0\\[2pt]
a_{21}b_{11}+a_{22}b_{21} & a_{21}b_{12}+a_{22}b_{22} & a_{21}b_{13}+a_{22}b_{23} & 0\\[2pt]
0 & 0 & 0 & 0\\[2pt]
0 & 0 & 0 & 0
\end{bmatrix}
}
\]

Finally, by discarding the padded zeros, we recover the actual product (for demonstration):
\[
\mathbf{A} \cdot \mathbf{B} =
\begin{bmatrix}
a_{11}b_{11} + a_{12}b_{21} & a_{11}b_{12} + a_{12}b_{22} & a_{11}b_{13} + a_{12}b_{23}\\[6pt]
a_{21}b_{11} + a_{22}b_{21} & a_{21}b_{12} + a_{22}b_{22} & a_{21}b_{13} + a_{22}b_{23}
\end{bmatrix}.
\]

Note that although this multiplication could be performed more efficiently in this local example, the padding is necessary to ensure proper alignment across the entire network. This padding is crucial for two reasons: (i) It allows for rotations that would deem some network configurations impossible due to insufficient free slots in the ciphertext. This happens when elements are packed together too closely, or when non-power-of-2 dimension representations overwrite valid values (ii) It aligns the rows and columns of operand matrices for more efficient processing, reducing the number of costly ciphertext rotation operations during training. 

\subsubsection{Dynamic Packing for Deep Networks.}
\label{sec:novel_global_pad}

The previous representation example was given with a local scope. For neural networks with multiple layers, additional challenges arise.
For example, if the amount of free slots required in the ciphertext is not calculated with a global view, there could be misalignment in further layers. Additionally, the representation created in the forward pass should be in a format that additional time is not wasted in the backward pass. 

Similar to the local-scope approach, we employ a dynamic packing scheme that adaptively switches layer weight matrices between row-major and column-major formats, following a strategy akin to the one proposed in~\cite{sav2021poseidon}. Odd-numbered layers use column-major packing, while even-numbered layers use row-major packing, allowing for more efficient operations throughout the network. By taking subsequent layer dimensions throughout the network into consideration, we calculate the multiplication area $\mu$ for each layer. 

\input{padding_algo_forward}

\begin{definition}[Dimension Slots Vector]
The dimension slots vector $\mathbf{d} = [d_0, d_1, d_2, \ldots, d_L]$ represents the padded dimensions for each layer, where:
\begin{itemize}
\item $d_0$ = padded input dimension
\item $d_i$ = padded dimension between layer $i-1$ and layer $i$
\item $d_L$ = padded output dimension
\end{itemize}
These are determined by the network configuration and rounded up to the nearest power of two.
\end{definition}

\begin{definition}[Maximum Slot Capacity]
The maximum slot capacity $\mathcal{N/2}$ represents the total number of slots available in a single ciphertext. See Section~\ref{sec:MHEbackground} for more information.
\end{definition}

\begin{definition}[Multiplication Area Vector]
The multiplication area vector $\boldsymbol{\mu} = [\mu_0, \mu_1, \mu_2, \ldots, \mu_{L-1}]$ defines how many slots are used for each layer, where $\mu_i \in \{1, 2, 4, 8, \ldots\}$ (powers of 2). $\mu_i$ determines the padding required for layer $i$, and is calculated via:
\end{definition}
\begin{align*}
\boldsymbol{\mu}[i] &= \frac{\mathcal{N}/2}{\mathbf{d}[i]\times \mathbf{d}[i+1]}
\end{align*}

Algorithms \ref{algo:layerhe-forward} and \ref{algo:layerhe-backward} detail our packing scheme and other network details for forward and backward passes, respectively. The forward pass algorithm initiates by selecting layer-index-dependent parameters using the input dimension slots vector $\mathbf{d}$ and $\boldsymbol{\mu}$ (Lines 2-9). Specifically, for odd-indexed layers, the padding step $s_{\text{pad}}$ (which determines the amount of padding between rows or columns) is determined by the current layer's multiplication area $\mu_r$ and the number of rows in the weight matrix, with column-wise padding $\pi_{\text{col}}$ and rotation parameters $\boldsymbol{\rho}_{\text{forward}} $ set using the dimension slot $\mathbf{d}[r-1]$ (Lines 3-5). Conversely, for even-indexed layers, $s_{\text{pad}}$ is based on $\mu_r$ and the matrix's column count, utilizing row-wise padding $\pi_{\text{row}}$ and rotation parameters derived from both $s_{\text{pad}}$ and $\mathbf{d}[r-1]$ (Lines 6-8). Following this, the weight matrix $\mathbf{W}$ is padded in plaintext space according to these parameters (Line 10). The linear transformation then occurs, producing the output $\mathbf{Z}_r$ by leveraging the input tensor, the padded weights, and the established rotation parameters $\boldsymbol{\rho}_{\text{forward}} $ (Line 12). If the current layer is not the final one, an activation function is applied to $\mathbf{Z}_r$ (Line 15). Subsequently, for odd-indexed layers, $\mathbf{Z}_r$ is rotated based on the dimension slot $\mathbf{d}[r+1]$ of the next layer (Line 18). For even-indexed layers, the rotation involves determining the next layer's dimensions $n_r, n_c$ from $\mathbf{d}$ (Line 20), retrieving the subsequent layer's slot multiplier $\mu_{r+1}$ from $\boldsymbol{\mu}[r]$ (Line 21), and then adjusting either $n_r$ or $n_c$ based on $\mu_{r+1}$ and the index of the next layer before the final rotation is applied to $\mathbf{Z}_r$ (Lines 22-27). This rotation is crucial for the alignment of subsequent layers and gradient calculation.

\input{padding_algo_backward}

The backward pass (Algorithm \ref{algo:layerhe-backward}) begins by applying the derivative of the activation function to the propagated error $\delta_r$, using the pre-activation output $\tilde{\mathbf Z}_r$, if the current layer is not the final one (Lines 2-4). For error-gradient preparation (Line 5), when the layer index $r$ is odd, we rotate $\delta_r$ along the dimension slot exactly $\mathbf{d}[r-1]$ times, using increasing successive offsets $1, 2, 4, 8,\dots$ up to the $\mathbf{d}[r-1]^{\text{th}}$ rotation. If the index is even, the rotation amount for $\delta_r$ is determined by dividing the maximum slot capacity $\mathcal{N}/2$ by $\mathbf{d}[r-1]$ (Line 8). Next, the weight gradient $\nabla W_r$ is computed via multiplication of the previous layer's output $\mathbf{Z}_{r-1}$ and the prepared $\delta_r$ (Line 11). The output gradient is propagated for all layers except the input layer. Layer-index-dependent rotation parameters $\boldsymbol{\rho}_{\text{backward}}$ are selected: for even indices, using a fixed rotation and $\mathbf{d}[r]$ (Line 16), and for odd indices, using $\frac{M}{\mathbf{d}[r]}$ (Line 18). Then, $\nabla\mathbf{W}_r$ is updated by multiplying $\delta_r$ with the transpose of the current layer's weights $\mathbf{W}_r$ using these rotation parameters (Line 20). Finally, during the weight update phase (Line 23), the change in weights $\Delta\mathbf{W}_r$ is calculated using the learning rate $\eta$ and $\nabla W_r$ (Line 24). The layer weights $\mathbf{W}_r$ are then updated by subtracting $\Delta\mathbf{W}_r$ (Line 26) and bootstrapped if necessary (Line 27). We monitor the ciphertext levels of the weight matrices, and if they approach a critical threshold, we perform a bootstrapping operation to refresh the noise before it becomes irrecoverably high. The algorithm concludes by returning the propagated gradient $\nabla\mathbf{W}_r$.

%% file: glossary.tex
\begin{table*}[h]
\centering
\caption{Glossary}
\label{Glossary}
\begin{tabular}{@{}l p{10cm}@{}}
\toprule
Symbol / Term & Description \\ \midrule
$\alpha$ & Fusion weight between encrypted and plaintext logits \\
$B$ & Mini\mbox{-}batch size \\
$b$ & A single batch inside a local epoch \\
BCO / BCD & Breast Cancer Wisconsin (Original / Diagnostic) datasets \\
CKKS & Cheon–Kim–Kim–Song homomorphic-encryption scheme \\
DMF & Deep Model Fusion \\
$d_i$ & Slot-padded dimension of layer $i$ \\
DP & Differential Privacy \\
FL & Federated Learning \\
$\mathcal{F}$ & Full feature set \\
$\mathcal{F}_{HE},\;\mathcal{F}_{P}$ & Features processed under HE / plaintext \\
GI & Global iteration (one server aggregation round) \\
GI-Load & Relative iterations per fixed data volume \\
HE & Homomorphic Encryption \\
\sys & Proposed framework \\
iDLG & Improved Deep Leakage from Gradients attack \\
$K$ & Total number of clients \\
$\mathbf{L}$ & Loss value \\
$L$ & Multiplicative depth (CKKS levels) \\
$n_k$ & Samples on client $k$ \\
$n_t$ & Total samples in round $t$ ($\sum_{k\in K} n_k$) \\
$\mathcal{N}$ & CKKS ring degree \\
PCA & Principal Component Analysis \\
$\pi_{\text{row}},\;\pi_{\text{col}}$ & Row/column flattening functions \\
$r$ & Layer index \\
$t$ & Training round index \\
$\rho$ & Rotation-parameter set \\
$\mu$ & Multiplication-area factor \\
MHE & Multiparty Homomorphic Encryption \\
$s_{\text{pad}}$ & Padding stride \\
$W_{HE}^{k,t},\;W_{P}^{k,t}$ & Client $k$ weights (HE / P) at round $t$ \\
$\nabla W_{HE}^{k,t},\;\nabla W_{P}^{k,t}$ & Gradients for HE / P sub-nets (client $k$, round $t$) \\
$\mathbf{z}_{HE},\;\mathbf{z}_{P},\;\bar{\mathbf{z}}$ & Logits (HE, plaintext, fused) \\ \bottomrule
\end{tabular}
\end{table*}

%% file: padding_algo_forward.tex
\begin{algorithm}[t]
\small
\caption{Forward Pass with Alternating-Packing-Based Padding Strategy}
\label{algo:layerhe-forward}
\begin{algorithmic}[1]
\Require Layer index $r \in \mathbb{N}$, input tensor $\mathbf{X} \in \mathbb{R}^{m \times n}$, weight matrix $\mathbf{W}_r \in \mathbb{R}^{n \times p}$, multiplication area $\boldsymbol{\mu} \in \mathbb{N}^k$, dimension slots $\mathbf{d} \in \mathbb{N}^k$.

\If{$r \equiv 1 \pmod{2}$} \Comment{Select appropriate padding and rotation params.}
    \State $s_{\text{pad}} \gets \mu_r \cdot n$
    \State $\pi_{\text{pad}} \gets \pi_{\text{col}}$ \Comment{Padding function $\pi_{\text{pad}}$ is now using column-padding strategy}
    \State $\boldsymbol{\rho}_{\text{forward}} \gets (1, \mathbf{d}[r-1])$
\Else
    \State $s_{\text{pad}} \gets \mu_r \cdot p$
    \State $\pi_{\text{pad}} \gets \pi_{\text{row}}$ \Comment{Padding function $\pi_{\text{pad}}$ is now using column-padding strategy}
    \State $\boldsymbol{\rho}_{\text{forward}}  \gets (s_{\text{pad}}, \mathbf{d}[r-1])$
\EndIf

\State $\mathbf{W}_r \gets \text{Pad}(\mathbf{W}_r, \pi_{\text{pad}}, s_{\text{pad}})$ \\

\State $\mathbf{Z}_r \gets \text{HEMultiply}(\mathbf{X}, \mathbf{W}_{\text{pad}}, \boldsymbol{\rho}_{\text{forward}} )$ \Comment{Forward Pass}
\State $\tilde{\mathbf Z}_r \gets \mathbf{Z}_r$ \Comment{Retain pre-activation output} \\

\If{$r \neq |\mathbf{d}| - 1$}
    \State $\mathbf{Z}_r \gets \sigma(\mathbf{Z}_r)$ \Comment{Apply activation function}
    
    \If{$r \equiv 1 \pmod{2}$}
        \State $\mathbf{Z}_r \gets \text{Rotate}(\mathbf{Z}_r, \text{right}, 1, \mathbf{d}[r+1])$
    \Else
        \State $n_r \gets \mathbf{d}[r], \quad n_c \gets \mathbf{d}[r+1]$
        \State $\mu_{r+1} \gets \boldsymbol{\mu}[r]$
        
        \If{$(r+1) \equiv 1 \pmod{2}$}
            \State $n_r \gets n_r \cdot \mu_{r+1}$
        \Else
            \State $n_c \gets n_c \cdot \mu_{r+1}$
        \EndIf
        
        \State $\mathbf{Z}_r \gets \text{Rotate}(\mathbf{Z}_r, \text{right}, n_r, n_c)$
    \EndIf
\EndIf

\State \Return $\mathbf{Z}_r$
\end{algorithmic}
\end{algorithm}

%% file: padding_algo_backward.tex
\begin{algorithm}[t]
\small
\caption{Backward Pass with Alternating-Packing-Based Gradient Processing}
\label{algo:layerhe-backward}
\begin{algorithmic}[1]
\Require Layer index $r$, input $\mathbf{Z}_{r-1}$, incoming error $\delta_r$, multiplication area $\boldsymbol{\mu}$, weights $\mathbf{W}_r$, maximum slot capacity $\mathcal{N}/2$

\If{$r \neq |\mathbf{d}| - 1$}
    \State $\delta_r \gets \delta_r \odot \sigma(\tilde{\mathbf Z}_r)$ \Comment{Activation derivative for pre-activation $\tilde{\mathbf Z}_r$}
\EndIf \\

\Statex \textit{\underline{Error Gradient Preparation:}}
\If{$r \equiv 1 \pmod{2}$} 
    \State $\delta_r \gets \text{Rotate}(\delta_r, \text{right}, 1, \mathbf{d}[r-1])$
\Else 
    \State $\delta_r\gets \text{Rotate}(\delta_r, \text{right}, \frac{\mathcal{N}/2}{\mathbf{d}[r-1]}, \mathbf{d}[r-1])$
\EndIf \\

\Statex \textit{\underline{Weight Gradient Computation:}}
\State $\nabla{W_r} \gets \text{HEMultiply}(\mathbf{Z}_{r-1}, \delta_r)$ \\

\Statex \textit{\underline{Output Gradient Propagation:}}
\State $\delta_{r-1} \gets \delta_r$
\If{$r > 1$}
    \If{$r \equiv 0 \pmod{2}$} 
        \State $\boldsymbol{\rho}_{\text{backward}} \gets (1, \mathbf{d}[r])$
    \Else 
        \State $\boldsymbol{\rho}_{\text{backward}} \gets (\frac{\mathcal{N}/2}{\mathbf{d}[r]}, \mathbf{d}[r])$
    \EndIf
    
    \State $\delta_{r-1} \gets \text{HEMultiply}(\delta_r, \mathbf{W}_r, \boldsymbol{\rho}_{\text{backward}})$
\EndIf
\State \Return $\delta_{r-1}, \,\nabla\mathbf{W}_r$ \\

\Statex \textit{\underline{Weight Update (After Aggregation):}}
\State $\Delta\mathbf{W}_r \gets \eta \cdot \nabla W_r$ \Comment{$\eta$ is learning rate, $\nabla W_r$ is now the aggregated layer gradient}

\State $\Delta\mathbf{W}_r \gets \text{Rotate}(\Delta\mathbf{W}, \text{right}, \mu_r \cdot \mathbf{d}[r], B)$

\State $\mathbf{W}_r \gets \mathbf{W}_r - \Delta\mathbf{W}_r$
\State $\mathbf{W}_r \gets \text{Bootstrap}(\mathbf{W}_r)$

\end{algorithmic}
\end{algorithm}

%% file: ref.bib
@article{ratra2022big,
  title={Big data privacy preservation using principal component analysis and random projection in healthcare},
  author={Ratra, Ritu and Gulia, Preeti and Gill, Nasib Singh and Chatterjee, Jyotir Moy},
  journal={Mathematical Problems in Engineering},
  volume={2022},
  number={1},
  pages={6402274},
  year={2022},
  publisher={Wiley Online Library}
}

@article{federatedLearning1,
  author    = {H. Brendan McMahan and
               Eider Moore and
               Daniel Ramage and
               Blaise Ag{\"{u}}era y Arcas},
  title     = {Federated Learning of Deep Networks using Model Averaging},
  journal   = {CoRR},
  volume    = {abs/1602.05629},
  year      = {2016},
  url       = {http://arxiv.org/abs/1602.05629},
  archivePrefix = {arXiv},
  eprint    = {1602.05629},
  bibsource = {dblp computer science bibliography, https://dblp.org}
}

@inproceedings{zhang2020gan,
  title={Gan enhanced membership inference: A passive local attack in federated learning},
  author={Zhang, Jingwen and Zhang, Jiale and Chen, Junjun and Yu, Shui},
  booktitle={IEEE International Conference on Communications (ICC)},
  pages={1--6},
  year={2020},
  organization={IEEE}
}

@inproceedings{Nasr2019,
author={M. {Nasr} and R. {Shokri} and A. {Houmansadr}},
title={Comprehensive Privacy Analysis of Deep Learning: Passive and Active White-box Inference Attacks against Centralized and Federated Learning}, 
booktitle={IEEE S\&P}, 
year={2019},
}

@ARTICLE{9160866,
  author={Al Badawi, Ahmad and Jin, Chao and Lin, Jie and Mun, Chan Fook and Jie, Sim Jun and Tan, Benjamin Hong Meng and Nan, Xiao and Aung, Khin Mi Mi and Chandrasekhar, Vijay Ramaseshan},
  journal={IEEE Transactions on Emerging Topics in Computing}, 
  title={Towards the AlexNet Moment for Homomorphic Encryption: HCNN, the First Homomorphic CNN on Encrypted Data With GPUs}, 
  year={2021},
  volume={9},
  number={3},
  pages={1330-1343},
  doi={10.1109/TETC.2020.3014636}}

@ARTICLE{8260844,
  author={Sun, Xiaoqiang and Zhang, Peng and Liu, Joseph K. and Yu, Jianping and Xie, Weixin},
  journal={IEEE Transactions on Emerging Topics in Computing}, 
  title={Private Machine Learning Classification Based on Fully Homomorphic Encryption}, 
  year={2020},
  volume={8},
  number={2},
  pages={352-364},
  doi={10.1109/TETC.2018.2794611}}

@INPROCEEDINGS{Melis2019,
  author={L. {Melis} and C. {Song} and E. {De Cristofaro} and V. {Shmatikov}},
  booktitle={2019 IEEE Symposium on Security and Privacy (SP)}, 
  title={Exploiting Unintended Feature Leakage in Collaborative Learning}, 
  year={2019},
  volume={},
  number={},
  pages={691-706},}

@inproceedings{Hitaj2017,
author = {Hitaj, Briland and Ateniese, Giuseppe and Perez-Cruz, Fernando},
title = {Deep Models Under the {GAN}: Information Leakage from Collaborative Deep Learning},
booktitle = {ACM CCS},
year = {2017},
}

@INPROCEEDINGS{9833648,
  author={Tian, Han and Zeng, Chaoliang and Ren, Zhenghang and Chai, Di and Zhang, Junxue and Chen, Kai and Yang, Qiang},
  booktitle={IEEE Symposium on Security and Privacy (SP)}, 
  title={Sphinx: Enabling Privacy-Preserving Online Learning over the Cloud}, 
  year={2022},
  volume={},
  number={},
  pages={2487-2501},
  doi={10.1109/SP46214.2022.9833648}}

@ARTICLE{10506637,
  author={Hu, Chenghao and Li, Baochun},
  journal={IEEE Transactions on Dependable and Secure Computing}, 
  title={MaskCrypt: Federated Learning With Selective Homomorphic Encryption}, 
  year={2025},
  volume={22},
  number={1},
  pages={221-233},
  doi={10.1109/TDSC.2024.3392424}}

@ARTICLE{hercules,
  author={Xu, Guowen and Han, Xingshuo and Xu, Shengmin and Zhang, Tianwei and Li, Hongwei and Huang, Xinyi and Deng, Robert H.},
  journal={IEEE Transactions on Dependable and Secure Computing}, 
  title={Hercules: Boosting the Performance of Privacy-Preserving Federated Learning}, 
  year={2023},
  volume={20},
  number={5},
  pages={4418-4433},
  doi={10.1109/TDSC.2022.3218793}}

@INPROCEEDINGS{Wang2019,
  author={Z. {Wang} and M. {Song} and Z. {Zhang} and Y. {Song} and Q. {Wang} and H. {Qi}},
  booktitle={IEEE INFOCOM 2019 - IEEE Conference on Computer Communications}, 
  title={Beyond Inferring Class Representatives: User-Level Privacy Leakage From Federated Learning}, 
  year={2019},
  volume={},
  number={},
  pages={2512-2520},}

@article{Konency2016fed,
  author={Kone{\v{c}}n{\`y}, Jakub and McMahan, H Brendan and Ramage, Daniel and Richt{\'a}rik, Peter},
  title={Federated optimization: Distributed machine learning for on-device intelligence},
  journal={CoRR},
  volume={abs:1610.02527},
  year={2016},
}

@article{Sav2022PrivacyPreservingFR,
  title={Privacy-Preserving Federated Recurrent Neural Networks},
  author={Sinem Sav and Abdulrahman Diaa and Apostolos Pyrgelis and Jean-Philippe Bossuat and Jean-Pierre Hubaux},
  journal={PoPETs},
  volume={2023},
  pages={500-521},
  url={https://api.semanticscholar.org/CorpusID:251135050}
}

@inproceedings{sav2021poseidon,
      title={POSEIDON: Privacy-Preserving Federated Neural Network Learning}, 
      author={Sinem Sav and Apostolos Pyrgelis and Juan R. Troncoso-Pastoriza and David Froelicher and Jean-Philippe Bossuat and Joao Sa Sousa and Jean-Pierre Hubaux},
      year={2021},
     booktitle={NDSS}
}

@inproceedings{shokri2015privacy,
  title={Privacy-preserving deep learning},
  author={Shokri, Reza and Shmatikov, Vitaly},
  booktitle={ACM Conference on Computer and Communications Security (CCS)},
  year={2015}
}

@inproceedings{zhang2020batchcrypt,
  title={$\{$BatchCrypt$\}$: Efficient homomorphic encryption for $\{$Cross-Silo$\}$ federated learning},
  author={Zhang, Chengliang and Li, Suyi and Xia, Junzhe and Wang, Wei and Yan, Feng and Liu, Yang},
  booktitle={2020 USENIX annual technical conference (ATC 20)},
  year={2020}
}

@inproceedings{bonawitz2017practical,
author = {Bonawitz, Keith and Ivanov, Vladimir and Kreuter, Ben and Marcedone, Antonio and McMahan, H. Brendan and Patel, Sarvar and Ramage, Daniel and Segal, Aaron and Seth, Karn},
title = {Practical Secure Aggregation for Privacy-Preserving Machine Learning},
year = {2017},
isbn = {9781450349468},
publisher = {Association for Computing Machinery},
address = {New York, NY, USA},
url = {https://doi.org/10.1145/3133956.3133982},
doi = {10.1145/3133956.3133982},
booktitle = {Proceedings of the 2017 ACM SIGSAC Conference on Computer and Communications Security},
pages = {1175–1191},
numpages = {17},
location = {Dallas, Texas, USA},
series = {CCS '17}
}

@inproceedings{hosseini2021secure,
  title={Secure aggregation in federated learning via multiparty homomorphic encryption},
  author={Hosseini, Erfan and Khisti, Ashish},
  booktitle={2021 IEEE Globecom Workshops (GC Wkshps)},
  pages={1--6},
  year={2021},
  organization={IEEE}
}

@article{mansouri2023sok,
  title={Sok: Secure aggregation based on cryptographic schemes for federated learning},
  author={Mansouri, Mohamad and {\"O}nen, Melek and Jaballah, Wafa Ben and Conti, Mauro},
  journal={PoPETs},
  year={2023}
}

@inproceedings{truex2020ldp,
  title={LDP-Fed: Federated learning with local differential privacy},
  author={Truex, Stacey and Liu, Ling and Chow, Ka-Ho and Gursoy, Mehmet Emre and Wei, Wenqi},
  booktitle={Proceedings of the third ACM international workshop on edge systems, analytics and networking},
  pages={61--66},
  year={2020}
}

@inproceedings{McMahan2018,
title={Learning Differentially Private Recurrent Language Models},
author={H. Brendan McMahan and Daniel Ramage and Kunal Talwar and Li Zhang},
booktitle={International Conference on Learning Representations},
year={2018},
url={https://openreview.net/forum?id=BJ0hF1Z0b},
}

@article{gao2020survey,
    author = {Gao, Jing and Li, Peng and Chen, Zhikui and Zhang, Jianing},
    title = {A Survey on Deep Learning for Multimodal Data Fusion},
    journal = {Neural Computation},
    volume = {32},
    number = {5},
    pages = {829-864},
    year = {2020},
    month = {05},
    issn = {0899-7667},
    doi = {10.1162/neco_a_01273},
    url = {https://doi.org/10.1162/neco\_a\_01273},
    eprint = {https://direct.mit.edu/neco/article-pdf/32/5/829/1865303/neco\_a\_01273.pdf},
}

@article{park2022privacy,
  title={Privacy-preserving federated learning using homomorphic encryption},
  author={Park, Jaehyoung and Lim, Hyuk},
  journal={Applied Sciences},
  volume={12},
  number={2},
  pages={734},
  year={2022},
  publisher={MDPI}
}

@article{baldiPCA1989,
title = {Neural networks and principal component analysis: Learning from examples without local minima},
journal = {Neural Networks},
volume = {2},
number = {1},
pages = {53-58},
year = {1989},
issn = {0893-6080},
doi = {https://doi.org/10.1016/0893-6080(89)90014-2},
url = {https://www.sciencedirect.com/science/article/pii/0893608089900142},
author = {Pierre Baldi and Kurt Hornik}
}

@INPROCEEDINGS{SP-PCA,
  author={Froelicher, David and Cho, Hyunghoon and Edupalli, Manaswitha and Sa Sousa, Joao and Bossuat, Jean-Philippe and Pyrgelis, Apostolos and Troncoso-Pastoriza, Juan R. and Berger, Bonnie and Hubaux, Jean-Pierre},
  booktitle={2023 IEEE Symposium on Security and Privacy (SP)}, 
  title={Scalable and Privacy-Preserving Federated Principal Component Analysis}, 
  year={2023},
  volume={},
  number={},
  pages={1908-1925},
  doi={10.1109/SP46215.2023.10179350}}

@article{PPPCA,
  title={Privacy Preserving PCA for Multiparty Modeling},
  author={Yingting Liu and Chaochao Chen and Longfei Zheng and L. xilinx Wang and Jun Zhou and Gui-Jie Liu},
  journal={ArXiv},
  year={2020},
  volume={abs/2002.02091},
  url={https://api.semanticscholar.org/CorpusID:211043587}
}

@phdthesis{mouchet2023multiparty,
  title={Multiparty homomorphic encryption: From theory to practice},
  author={Mouchet, Christian Vincent},
  year={2023},
  school={EPFL}
}

@article{mouchet2021multiparty,
  title={Multiparty homomorphic encryption from ring-learning-with-errors},
  author={Mouchet, Christian and Troncoso-Pastoriza, Juan and Bossuat, Jean-Philippe and Hubaux, Jean-Pierre},
  journal={PoPETs},
  volume={2021},
  number={4},
  pages={291--311},
  year={2021}
}

@misc{zhao2020idlgimproveddeepleakage,
      title={iDLG: Improved Deep Leakage from Gradients}, 
      author={Bo Zhao and Konda Reddy Mopuri and Hakan Bilen},
      year={2020},
      eprint={2001.02610},
      archivePrefix={arXiv},
      primaryClass={cs.LG},
      url={https://arxiv.org/abs/2001.02610}, 
}

@Article{         harris2020array,
 title         = {Array programming with {NumPy}},
 author        = {Charles R. Harris and K. Jarrod Millman and St{\'{e}}fan J.
                 van der Walt and Ralf Gommers and Pauli Virtanen and David
                 Cournapeau and Eric Wieser and Julian Taylor and Sebastian
                 Berg and Nathaniel J. Smith and Robert Kern and Matti Picus
                 and Stephan Hoyer and Marten H. van Kerkwijk and Matthew
                 Brett and Allan Haldane and Jaime Fern{\'{a}}ndez del
                 R{\'{i}}o and Mark Wiebe and Pearu Peterson and Pierre
                 G{\'{e}}rard-Marchant and Kevin Sheppard and Tyler Reddy and
                 Warren Weckesser and Hameer Abbasi and Christoph Gohlke and
                 Travis E. Oliphant},
 year          = {2020},
 month         = sep,
 journal       = {Nature},
 volume        = {585},
 number        = {7825},
 pages         = {357--362},
 doi           = {10.1038/s41586-020-2649-2},
 publisher     = {Springer Science and Business Media {LLC}},
 url           = {https://doi.org/10.1038/s41586-020-2649-2}
}

@ARTICLE{2020SciPy,
  author  = {Virtanen, Pauli and Gommers, Ralf and Oliphant, Travis E. and
            Haberland, Matt and Reddy, Tyler and Cournapeau, David and
            Burovski, Evgeni and Peterson, Pearu and Weckesser, Warren and
            Bright, Jonathan and {van der Walt}, St{\'e}fan J. and
            Brett, Matthew and Wilson, Joshua and Millman, K. Jarrod and
            Mayorov, Nikolay and Nelson, Andrew R. J. and Jones, Eric and
            Kern, Robert and Larson, Eric and Carey, C J and
            Polat, {\.I}lhan and Feng, Yu and Moore, Eric W. and
            {VanderPlas}, Jake and Laxalde, Denis and Perktold, Josef and
            Cimrman, Robert and Henriksen, Ian and Quintero, E. A. and
            Harris, Charles R. and Archibald, Anne M. and
            Ribeiro, Ant{\^o}nio H. and Pedregosa, Fabian and
            {van Mulbregt}, Paul and {SciPy 1.0 Contributors}},
  title   = {{{SciPy} 1.0: Fundamental Algorithms for Scientific
            Computing in Python}},
  journal = {Nature Methods},
  year    = {2020},
  volume  = {17},
  pages   = {261--272},
  adsurl  = {https://rdcu.be/b08Wh},
  doi     = {10.1038/s41592-019-0686-2},
}

@book{R._2000, title={Practical Methods of Optimization}, publisher={Wiley}, author={R., Fletcher}, year={2000} }

@article{you2024local,
  title={Local Differential Privacy is Not Enough: A Sample Reconstruction Attack against Federated Learning with Local Differential Privacy},
  author={You, Zhichao and Dong, Xuewen and Li, Shujun and Ma, Siqi and Shen, Yulong},
  journal={IEEE Transactions on Information Forensics and Security},
  year={2024},
  publisher={IEEE}
}

@article{bagdasaryan2020backdoor,
  author       = {Eugene Bagdasaryan and
                  Andreas Veit and
                  Yiqing Hua and
                  Deborah Estrin and
                  Vitaly Shmatikov},
  title        = {How To Backdoor Federated Learning},
  journal      = {CoRR},
  volume       = {abs/1807.00459},
  year         = {2018},
  url          = {http://arxiv.org/abs/1807.00459},
  eprinttype    = {arXiv},
  eprint       = {1807.00459},
  bibsource    = {dblp computer science bibliography, https://dblp.org}
}

@InProceedings{bachrach16,
  title = 	 {CryptoNets: Applying Neural Networks to Encrypted Data with High Throughput and Accuracy},
  author = 	 {Gilad-Bachrach, Ran and Dowlin, Nathan and Laine, Kim and Lauter, Kristin and Naehrig, Michael and Wernsing, John},
  booktitle = 	 {Proceedings of The 33rd International Conference on Machine Learning},
  pages = 	 {201--210},
  year = 	 {2016},
  editor = 	 {Balcan, Maria Florina and Weinberger, Kilian Q.},
  volume = 	 {48},
  series = 	 {Proceedings of Machine Learning Research},
  address = 	 {New York, New York, USA},
  month = 	 {20--22 Jun},
  publisher =    {PMLR},
  url = 	 {https://proceedings.mlr.press/v48/gilad-bachrach16.html}
}

@InProceedings{Nandakumar_2019_CVPR_Workshops,
author = {Nandakumar, Karthik and Ratha, Nalini and Pankanti, Sharath and Halevi, Shai},
title = {Towards Deep Neural Network Training on Encrypted Data},
booktitle = {Proceedings of the IEEE/CVF Conference on Computer Vision and Pattern Recognition (CVPR) Workshops},
month = {June},
year = {2019}
}

@article{gronberg2025blindfl,
  title={BlindFL: Segmented Federated Learning with Fully Homomorphic Encryption},
  author={Gronberg, Evan and d'Aliberti, Liv and Saebo, Magnus and Hook, Aurora},
  journal={arXiv preprint arXiv:2501.11659},
  year={2025}
}

@misc{OpenFHE,
      author = {Ahmad Al Badawi and Andreea Alexandru and Jack Bates and Flavio Bergamaschi and David Bruce Cousins and Saroja Erabelli and Nicholas Genise and Shai Halevi and Hamish Hunt and Andrey Kim and Yongwoo Lee and Zeyu Liu and Daniele Micciancio and Carlo Pascoe and Yuriy Polyakov and Ian Quah and Saraswathy R.V. and Kurt Rohloff and Jonathan Saylor and Dmitriy Suponitsky and Matthew Triplett and Vinod Vaikuntanathan and Vincent Zucca},
      title = {{OpenFHE}: Open-Source Fully Homomorphic Encryption Library},
      howpublished = {Cryptology ePrint Archive, Paper 2022/915},
      year = {2022},
      note = {\url{https://eprint.iacr.org/2022/915}},
      url = {https://eprint.iacr.org/2022/915}
}

@inproceedings{Netzer2011ReadingDI,
  title={Reading Digits in Natural Images with Unsupervised Feature Learning},
  author={Yuval Netzer and Tao Wang and Adam Coates and A. Bissacco and Bo Wu and A. Ng},
  year={2011},
  url={https://api.semanticscholar.org/CorpusID:16852518}
}

@article{lecun-mnisthandwrittendigit-2010,
  author = {LeCun, Yann and Cortes, Corinna},
  howpublished = {http://yann.lecun.com/exdb/mnist/},
  lastchecked = {2016-01-14 14:24:11},
  title = {{MNIST} handwritten digit database},
  url = {http://yann.lecun.com/exdb/mnist/},
  username = {mhwombat},
  year = 2010
}

@article{lu2023privacy,
author = {Lu, Yang and Yu, Zhengxin and Suri, Neeraj},
title = {Privacy-preserving Decentralized Federated Learning over Time-varying Communication Graph},
year = {2023},
issue_date = {August 2023},
publisher = {Association for Computing Machinery},
address = {New York, NY, USA},
volume = {26},
number = {3},
issn = {2471-2566},
url = {https://doi.org/10.1145/3591354},
doi = {10.1145/3591354},
journal = {ACM Trans. Priv. Secur.},
month = jun,
articleno = {33},
numpages = {39}
}

@ARTICLE{ssim,
  author={Zhou Wang and Bovik, A.C. and Sheikh, H.R. and Simoncelli, E.P.},
  journal={IEEE Transactions on Image Processing}, 
  title={Image quality assessment: from error visibility to structural similarity}, 
  year={2004},
  volume={13},
  number={4},
  pages={600-612},
  doi={10.1109/TIP.2003.819861}}

@INPROCEEDINGS{lpips,
  author={Zhang, Richard and Isola, Phillip and Efros, Alexei A. and Shechtman, Eli and Wang, Oliver},
  booktitle={2018 IEEE/CVF Conference on Computer Vision and Pattern Recognition}, 
  title={The Unreasonable Effectiveness of Deep Features as a Perceptual Metric}, 
  year={2018},
  volume={},
  number={},
  pages={586-595},
  doi={10.1109/CVPR.2018.00068}}
